\documentclass[10pt,journal,a4paper,twocolumn, compsoc]{IEEEtran}
\usepackage{graphicx}
\usepackage{amsmath}
\usepackage{amssymb}
\usepackage{multirow}
\usepackage{url}
\usepackage{capt-of}
\usepackage{cite}

\usepackage{booktabs}

\newcommand{\eg}{e.\,g.\ }
\newcommand{\ie}{i.\,e.\ }

\newtheorem{definition}{Definition}
\newtheorem{lemma}{Lemma}
\newtheorem{theorem}{Theorem}

\newcommand{\mimp}{\textsf{MIMP}}

\title{The SCJ Small Parsimony Problem for Weighted Gene Adjacencies}
\author{Nina Luhmann, Manuel Lafond, Annelyse Th\'{e}venin, A\"{\i}da Ouangraoua, Roland Wittler and Cedric Chauve

\IEEEcompsocitemizethanks{\IEEEcompsocthanksitem N. Luhmann and R. Wittler are with the International Research Training Group ``Computational Methods for the Analysis of the Diversity and Dynamics of Genomes'', Bielefeld University, Germany.
 \IEEEcompsocthanksitem N. Luhmann, R. Wittler and A. Th\'{e}venin are with the Genome Informatics group, Faculty of Technology and Center for Biotechnology, Bielefeld University, Germany.
  \IEEEcompsocthanksitem M. Lafond is with the Department of Computer Science and Operational Research, Universit\'{e} de Montr\'{e}al, Montr\'{e}al, Canada.
  \IEEEcompsocthanksitem A. Ouangraoua is with the Department of Computer Science, Universit\'{e} de Sherbrooke, Sherbrooke, Canada.
  \IEEEcompsocthanksitem C. Chauve is with the Department of Mathematics, Simon Fraser University, Burnaby (BC), Canada.}}

\begin{document}
%
\IEEEtitleabstractindextext{%
\begin{abstract}
Reconstructing ancestral gene orders in a given phylogeny is a classical problem in comparative genomics. Most existing methods compare conserved features in extant genomes in the phylogeny to define potential ancestral gene adjacencies, and either try to reconstruct all ancestral genomes under a global evolutionary parsimony criterion, or, focusing on a single ancestral genome, use a scaffolding approach to select a subset of ancestral gene adjacencies, generally aiming at reducing the fragmentation of the reconstructed ancestral genome.

In this paper, we describe an exact algorithm for the Small Parsimony Problem that combines both approaches. We consider that gene adjacencies at internal nodes of the species phylogeny are weighted, and we introduce an objective function defined as a convex combination of these weights and the evolutionary cost under the Single-Cut-or-Join (SCJ) model. 
The weights of ancestral gene adjacencies can \eg be obtained through the recent availability of ancient DNA sequencing data, which provide a direct hint at the genome structure of the considered ancestor, or through probabilistic analysis of gene adjacencies evolution. We show the NP-hardness of our problem variant and propose a Fixed-Parameter Tractable algorithm based on the Sankoff-Rousseau dynamic programming algorithm that also allows to sample co-optimal solutions. 
We apply our approach to mammalian and bacterial data providing different degrees of complexity.
We show that including adjacency weights in the objective has a significant impact in reducing the
fragmentation of the reconstructed ancestral gene orders. \\
An implementation is available at \url{http://github.com/nluhmann/PhySca}.
\end{abstract}}
\maketitle
%
%
\IEEEraisesectionheading{\section{Introduction}\label{sec:introduction}}
Reconstructing ancestral gene orders  is a long-standing computational biology problem with important applications, as shown in several recent large-scale projects~\cite{COFFEE,PINEAPPLE,ANOPHELES}. Informally, the problem can be defined as follows: Given a phylogenetic tree representing the speciation history leading to a set of extant genomes, we want to reconstruct the structure of the ancestral genomes corresponding to the internal nodes of the tree.

Existing ancestral genome reconstruction methods concentrate on two main strategies.  
\textit{Local} approaches consider the reconstruction of one specific ancestor at a time independently from the other ancestors of the tree. Usually, they do not consider an evolutionary model and  proceed in two stages: (1)  comparing gene orders of ingroup and outgroup species to define potential ancestral gene adjacencies, and (2) selecting a conflict-free subset of ancestral gene adjacencies  -- where a conflict is defined as an ancestral gene extremity belonging to more than two {potential} adjacencies, \eg due to convergent evolution --, 
to obtain a set of  Contiguous Ancestral Regions (CARs)~\cite{bertrand2010reconstruction,chauve2008methodological,ma2006reconstructing}. 
The second stage of this approach is often defined as a combinatorial optimization problem aiming to minimize the number 
of discarded ancestral adjacencies, thus maximizing the number of selected adjacencies ~\cite{bertrand2010reconstruction,ma2006reconstructing,mavnuch2012linearization}. This stage follows principles common in \textit{scaffolding} methods used to obtain gene orders for extant genomes from sequencing data~\cite{Bosi2015medusa,mandric2015scaffmatch}. This approach was recently used to scaffold an ancestral pathogen genome for which ancient DNA (aDNA) sequencing data could be obtained~\cite{rajaraman2013fpsac}.
\textit{Global} approaches on the other hand simultaneously reconstruct ancestral gene orders at all internal nodes of the considered phylogeny, generally based on a parsimony criterion within an evolutionary model. This so called \textit{Small Parsimony Problem} has been studied with several underlying genome rearrangement models, such as the breakpoint distance or the Double-Cut-and-Join (DCJ) distance~\cite{alekseyev2009breakpoint,kovac2011practical,zheng2011pathgroups}. While rearrangement scenarios based on complex rearrangement models can give insights into underlying evolutionary mechanisms, from a computational point of view, the Small Parsimony Problem is NP-hard for most rearrangement distances~\cite{tannier2009multichromosomal}. 
One exception is the Single-Cut-or-Join (SCJ) distance, for which linear/circular ancestral gene orders can be found in polynomial time~\cite{feijao2011scj}, however constraints required to ensure algorithmic tractability yield fragmented ancestral gene orders. 

The two approaches outlined above optimize somewhat orthogonal criteria. For example, the underlying goal of the local approach is to maximize the agreement between the resulting ancestral gene order and the set of potential ancestral adjacencies, independently of the other ancestral gene orders. Would it be applied independently to all ancestral nodes, potential ancestral adjacencies exhibiting a mixed profile of presence/absence in the extant genomes might then lead to a set of  non-parsimonious ancestral gene orders. 
The global approach aims only at minimizing the evolutionary cost in the phylogeny and can result in more fragmented ancestral gene orders. Nevertheless, there is little ground to claim that one approach or the other is more accurate or to be preferred, and the work we present is an attempt to reconcile both approaches. 

We introduce a variant of the Small Parsimony Problem based on an optimality criterion that accounts for both an evolutionary distance and the difference between the initial set of potential ancestral adjacencies and the final consistent subset of adjacencies conserved at each ancestral node. More precisely we consider that each potential ancestral gene adjacency can be provided with a (prior) non-negative weight at every internal node. 
The contribution of the discarded adjacencies to the objective function is then the sum of their weights. 
These adjacency weights can \eg be obtained as probabilities computed by sampling scenarios for each potential adjacency independently~\cite{chauve2014evolution} or can be based on ancient DNA (aDNA) sequencing data providing direct prior information assigned to certain ancestral nodes. It follows that the phylogenetic framework we present can then also assist in scaffolding fragmented assemblies of aDNA sequencing data~\cite{luhmann2014scaffolding,rajaraman2013fpsac}. 

We prove NP-hardness of the problem variant we introduce and 
describe an exact exponential time algorithm for reconstructing consistent ancestral genomes under this optimality criterion, based on a mixed Dynamic Programming / Integer Linear Programming approach.
We show that this Small Parsimony Problem variant is Fixed-Parameter Tractable (FPT), with a parameter linked to the amount of conflict in the data. Moreover, this also allows us to provide an FPT sampling algorithm for co-optimal solutions, -- a problem recently addressed in~\cite{miklos2015} using a MCMC approach.
We evaluate our method on a simulated dataset and compare our results to several other methods reconstructing ancestral genomes. Further, we apply our method to two real data sets: mammalian genomes spanning roughly one million years of evolution, and bacterial genomes (pathogen \textit{Yersinia}) spanning $20,000$ years of evolution and for which some aDNA sequencing data is available. We show that we can reduce the fragmentation of ancestral gene orders in both datasets by integrating adjacency weights while reconstructing robust ancestral genomes.

This paper is an extended version of the work previously presented in~\cite{luhmann2016scj}, particularly including new results on simulated datasets and a hardness proof of the defined problem.

%
%
\section{Background and problem statement}
\label{sec:background}

\subsection{Genomes and adjacencies}
Genomes consist of chromosomes and plasmids. Each such component can be represented as a linear or circular sequence of oriented markers over a marker alphabet. Markers correspond to homologous sequences between genomes, \eg genes or synteny blocks. We assume that each marker appears exactly once in each genome, so our model does not consider duplications or deletions. To account for its orientation, each marker~$x$ is encoded as a pair of marker extremities $(x_h,x_t)$ or $(x_t,x_h)$.

An \textit{adjacency} is an unordered pair of marker extremities, \eg $\{x_t,y_h\}$. The order of markers in a genome can be encoded by a set of adjacencies. Two distinct adjacencies are said to be \textit{conflicting} if they share a common marker extremity. If a set of adjacencies contains conflicting adjacencies, it is not \textit{consistent} with a mixed linear/circular genome model. 
We assume that the set of adjacencies for an extant assembled genome is consistent. The
set of adjacencies for one genome naturally defines an \textit{adjacency graph},
where nodes represent marker extremities and edges represent adjacencies.
Conflicting adjacencies can be identified as branching nodes in this graph.

\subsection{The Small Parsimony Problem and rearrangement distances}
In a global phylogenetic approach, we are given a phylogenetic tree with extant genomes at its leaves and internal nodes representing ancestral genomes. 
We denote by $\mathcal{A}$ the set of all adjacencies present in at least one extant genome and assume that every ancestral adjacency belongs to $\mathcal{A}$.
Then the goal is to find a labeling of the internal nodes by consistent subsets of $\mathcal{A}$ minimizing a chosen genomic distance over the tree. This is known as the \emph{Parsimonious Labeling Problem.}\\

\begin{definition}[Parsimonious Labeling Problem] 
\label{def:parsimony}
Let $T=(V,E)$ be a tree with each leaf $l$ labeled with a consistent set of adjacencies $\mathcal{A}_l \subseteq \mathcal{A}$, and $d$ a distance between consistent sets of adjacencies. A labeling $\lambda : V \rightarrow \mathcal{P}(\mathcal{A})$
with $\lambda(l) = \mathcal{A}_l$ for each leaf is parsimonious for $d$ if none of the internal nodes $v \in V$ contains a conflict and it minimizes the sum $W(\lambda,T)$ of the distances along the branches of $T$:
$$W(\lambda,T)=\sum_{(u,v) \in E} d\bigl(\lambda(u),\lambda(v)\bigr).$$
\end{definition}

This problem is NP-hard for most rearrangement distances taken as evolutionary models.
The only  known exception is the set-theoretic Single-Cut-or-Join (SCJ) distance~\cite{feijao2011scj}. It defines a rearrangement distance by two operations: the \textit{cut} and \textit{join} of adjacencies. Given two genomes defined by consistent sets of adjacencies $A$ and $B$, the SCJ distance between these genomes is $$d_{SCJ}(A,B)\;=\; \mid A-B\mid+\mid B-A\mid.$$

The Small Parsimony Problem under the SCJ model
can be solved by computing a parsimonious gain/loss history for each adjacency separately with the dynamic programming Fitch algorithm~\cite{fitch1971toward,HAR-1973} in polynomial time. 
Consistent labelings can be ensured with the additional constraint that in case of ambiguity at the root of the tree, the absence of the adjacency is chosen~\cite{feijao2011scj}. As each adjacency is treated independently, this constraint might automatically exclude all adjacencies being part of a conflict to ensure consistency.
This results in an unnecessarily sparse reconstruction  in terms of reconstructed adjacencies and thus more fragmented genomes higher up in the tree.

\subsection{Generalization by weighting adjacencies}

When considering an internal node~$v$, we define node~$u$ as its parent node in $T$. 
We assume that a specific adjacency graph is associated to each ancestral node~$v$, whose edges are annotated by a weight~$w_{v,a} \in [0,1]$ representing a confidence measure for the presence of adjacency~$a$ in species~$v$. 
Then in a global reconstruction, cutting an adjacency of a higher weight has higher impact in terms of the optimization criterion than cutting an adjacency of lower weight.

Formally, we define two additional variables for each adjacency $a \in \mathcal{A}$ at each internal node~$v \in V$:
The presence (or absence) of $a$ at node $v$ is represented by $p_{v,a} \in \{0,1\}$, while $c_{v,a} \in~\{0,1\}$ indicates a change for the status of an adjacency along an edge~$(u,v)$, i.e., $p_{u,a} \neq p_{v,a}$. 
We consider the problem of optimizing the following objective function, where $\alpha \in [0, 1]$ is a convex combination factor.\\

\begin{definition}[Weighted SCJ Labeling Problem]
\label{def:weightedParsimony}
Let $T=(V,E)$ be a tree with each leaf~$l$ labeled with a consistent set of adjacencies $\mathcal{A}_l \subseteq \mathcal{A}$ and each adjacency $a\in \mathcal{A}$ is assigned a given weight $w_{v,a}\in [0,1]$ for each node $v \in V$. A labeling $\lambda$ of the internal nodes of~$T$ with $\lambda(l) = \mathcal{A}_l$ for each leaf is an \emph{optimal weighted SCJ labeling} if none of the internal nodes $v \in V$ contains a conflict and it minimizes the criterion
\[D(\lambda,T)=\sum_{v,a} \alpha (1-p_{v,a}) w_{v,a} + (1-\alpha) c_{v,a}\]
\end{definition}

Further, we can state the corresponding co-optimal sampling problem. A sampling method is important to examine different co-optimal rearrangement scenarios that can explain evolution toward the structure of extant genomes. \\

\begin{definition}[Weighted SCJ Sampling Problem]
Given the setting of the Weighted SCJ Labeling Problem, sample uniformly from all labelings~$\lambda$ of the internal nodes of $T$ that are solutions to the Weighted SCJ Labeling Problem. 
\end{definition}

\subsection{Problem complexity}
Aside of the many heuristics for the Small Parsimony Problem for non-SCJ rearrangement models (see for example~\cite{xu2011gasts,zheng2011pathgroups,kovac2011practical} for the DCJ distance), 
there exist a few positive results for the Weighted SCJ Labeling Problem with specific values of $\alpha$. 

If $\alpha = 0$, the objective function corresponds to the Small Parsimony Problem under the SCJ distance and hence a solution can be found in polynomial time~\cite{feijao2011scj}. A generalization towards multifurcating, edge-weighted trees including prior information on adjacencies at exactly one internal node of the tree is given in~\cite{luhmann2014scaffolding}. 
Recently, Mikl\'{o}s and Smith~\cite{miklos2015} proposed a Gibbs sampler for sampling optimal labelings under the SCJ model with equal branch lengths. It starts from an optimal labeling obtained as in~\cite{feijao2011scj}, and then explores the space of co-optimal labelings through repeated constrained parsimonious modifications of a single adjacency evolutionary scenario. 
This method addresses the issue of the high fragmentation of internal node labelings, but convergence is not proven, and so there is no bound on the computation time.

If $\alpha = 1$, i.e., we do not take evolution in terms of SCJ distance along the branches of the tree into account, we can solve the problem by applying independently a maximum-weight matching
algorithm at each internal node~\cite{mavnuch2012linearization}. 
So the extreme cases of the problem are tractable, and while we assume that the general problem is hard, we will now prove it for a small range of $\alpha$.\\

\begin{theorem}\label{theorem:SCJ}
The Weighted SCJ Labeling Problem is NP-hard for any $\alpha > 33/34$.\\
\end{theorem}

We show the hardness by reduction from the Maximum Intersection Matching Problem, which is defined as follows.
Let $G_1$ and $G_2$ be two graphs on the same vertex set. Find a perfect matching in $G_1$ and $G_2$  such that the number of edges common to both matchings is maximized. We prove NP-hardness of this problem by reduction from 3-Balanced-Max-2-SAT (see appendix for details).\\

\begin{theorem}\label{theorem:MIMP}
The Maximum Intersection Matching Problem is NP-complete.\\
\end{theorem}

The relation of the Weighted SCJ Labeling Problem and the Maximum Intersection Matching Problem can be sketched as follows.
For a given instance of the Maximum Intersection Matching Problem, $G_1$ and $G_2$, we construct a tree that contains the edges of both graphs as potential adjacencies. For $\alpha > 33/34$, 
an optimal labeling of two internal nodes then corresponds to perfect matchings in $G_1$ and $G_2$. Maximizing the number of common edges of the matching then minimizes the SCJ distance between the nodes. A detailed proof is given in the appendix.

%
%
\section{Methods}
\label{sec:methods}

In order to find a solution to the Weighted SCJ Labeling Problem, we first show that we can decompose the problem into smaller independent subproblems. Then, for each subproblem containing conflicting adjacencies, we show that, if it contains a moderate level of conflict, it can be solved using the Sankoff-Rousseau algorithm~\cite{sankoff1975locating} with a  complexity parameterized by the size of the subproblem. For a highly conflicting subproblem, we show that it can be solved  by an Integer Linear Program (ILP).

\subsection{Decomposition into independent subproblems}
We first introduce a graph that encodes all adjacencies present in at least one internal node of the considered phylogeny (Def.~\ref{def:gag} and Supplementary Fig.~\ref{fig:gag}). As introduced previously, we consider a tree $T=(V,E)$ where each node is augmented with an adjacency graph.\\

\begin{definition}[Global adjacency graph]
\label{def:gag}
The set of vertices $V_{AG}$ of the global adjacency graph $AG$ consists of all marker extremities present in at least one of the adjacency graphs. There is an edge between two vertices $a, b\in V_{AG}$ that are not extremities of a same marker, if there is an internal node in the tree $T$ whose adjacency graph contains the adjacency $\{a,b\}$. The edge is labeled with the list of all internal nodes that contain this adjacency.\\
\end{definition}

Each connected component~$C$ of the global adjacency graph defines a subproblem composed of the species phylogeny, the set of marker extremities equal to the vertex set of $C$, and the set of adjacencies equal to the edge set of $C$. 
According to the following lemma, whose proof is straightforward, it is sufficient to solve each such subproblem independently.\\

\begin{lemma}
  \label{lem:subproblems}
  The set of all optimal solutions of the Weighted SCJ
  Labeling Problem is the set-theoretic Cartesian product of the sets
  of optimal solutions of the instances defined by the connected
  components of the global adjacency graph.\\
\end{lemma}

To solve the problem defined by a connected component~$C$ of the global adjacency graph containing conflicts, we rely on an adaptation of the Sankoff-Rousseau algorithm with exponential time complexity, parameterized by the size and nature of conflicts of $C$, and thus can solve subproblems with moderate amount of conflict.

\subsection{Overview of the Sankoff-Rousseau algorithm} The Sankoff-Rousseau dynamic programming algorithm~\cite{sankoff1975locating} solves the general Small Parsimony Problem for discrete characters. Let $L$ be the set of all possible labels of a node in the phylogeny. 
Then for each node~$u$ in the tree, the cost $c(a,u)$ of assigning a label $a \in L$ to this node is defined recursively as follows:
$$c(a,u)= \sum_{v \mbox{ \scriptsize child of } u} \min_{b \in L} \bigl(c(b,v) + d(a,b)\bigr)$$
where in our case $d(a,b)$ is defined as in Definition~\ref{def:weightedParsimony}.
This equation defines a dynamic programming algorithm whose base case is when $u$ is a leaf in which case $c(a,u)=0$ if $\lambda(u)=a$ and $c(a,u)=\infty$ otherwise. 
Afterwards, choosing a label with the minimum cost at the root node and backtracking in a top-down traversal of the tree results in a most parsimonious labeling. We refer to~\cite{csuros2013howto} for a review on the Sankoff-Rousseau algorithm.

\subsection{Application to the Weighted SCJ Labeling Problem} In order to use the Sankoff-Rousseau algorithm to solve the problem defined by a connected component~$C$ of the global adjacency graph, we define a label of an internal node of the phylogeny as the assignment of at most one adjacency to each marker extremity. More precisely, let $x$ be a marker extremity in $C$, $v$ an internal node of $T$, and $e_1,\dots,e_{d_x}$ be all edges in the global adjacency graph that are incident to $x$ and whose label contains $v$ (i.e., represent adjacencies in the adjacency graph of node $v$). We define the set of possible labels of $v$ as $L_{x,v}=\{\emptyset,e_1,\dots,e_{d_x}\}$. The set of potential labels~$L_v$ of node $v$ is then the Cartesian product of the label sets $L_{x,v}$ for all $x\in V(C)$ resulting in a set of discrete labels for $v$ of size $\prod_{x\in V(C)}(1+d_x)$. 
Note that not all of these joint labelings are valid as they can assign an adjacency $a=(x,y)$ to $x$ but not to $y$, or adjacency $a=(x,y)$ to $x$ and $b=(x,z)$ to $z$ thus creating a conflict (see Supplementary Fig.~\ref{fig:example} for an example).

For an edge $(u,v)$ in the tree, we can then define a cost matrix that is indexed by pairs of labels of $L_u$ and $L_v$, respectively. The cost is infinite if one of the labels is not valid, and defined by the objective function otherwise. We can then apply the Sankoff-Rousseau approach to find an optimal labeling of all internal nodes of the tree for connected component $C$. 

Note that, if $C$ is a connected component with no conflict, it is composed of two vertices and a single edge, and can be solved in space $O(n)$ and time $O(n)$.

\subsection{Complexity analysis} The time and space complexity of the algorithm is obviously exponential in the size of $C$. Indeed, the time (resp. space) complexity of the Sankoff-Rousseau algorithm for an instance with a tree having $n$ leaves and $r$ possible labels for each node is $O(nr^2)$ (resp. $O(nr)$)~\cite{csuros2013howto}. In our algorithm, assuming $n$ leaves in $T$ (\textit{i.e.} $n$ extant species), $m_C$ vertices in the global adjacency graph of $C$  and a maximum degree $d_C$ for vertices (marker extremities) in this graph, $(1+d_C)^{m_C}$ is an upper bound for the size of the label set~$L_v$ for a node~$v$.
Moreover, computing the distance between two labels of $L_v$ and $L_u$, where $(u,v)$ is an edge of $T$, can trivially be done in time and space $O(m_C)$: If both labels are valid, it suffices to check how many common adjacencies are present in both labels, while deciding if a label is not valid can be done by a one-pass examination of the label. Combining this with the Sankoff-Rousseau complexity yields a time complexity in $O(nm_C(1+d_C)^{2m_C})$ and a space complexity in $O(nm_C(1+d_C)^{m_C})$.

Given a general instance, i.e., an instance not limited to a single connected component of the global adjacency graph, we can consider each connected component independently (Lemma~\ref{lem:subproblems}). 
For a set of $N$ markers and $c$ connected components in the global adjacency graph defining a conflicting instance, we define $D$ as the maximum degree of a vertex and $M$ as the maximum number of vertices in all such components. Then, the complexity analysis above shows that the problem is Fixed-Parameter Tractable (FPT). \\

\begin{theorem}\label{thm:opt}
  The Weighted SCJ Labeling Problem can be solved in worst-case time $O(nN(1+D)^{2M})$ and space $O(nN(1+D)^{M})$.\\
\end{theorem}

In practice, the exponential complexity of our algorithm depends on the structure of the conflicting connected components of the global adjacency graph. The dynamic programming algorithm will be effective on instances with either small conflicting connected components or small degrees within such components, and will break down with a single component with a large number of vertices of high degree. For such components, the time complexity is provably high and we propose an ILP to solve them. 

\subsection{An Integer Linear Program}
We can formulate the optimization problem as a simple ILP. We consider two variables for any adjacency~$a$ and node~$v$, $p_{v,a} \in \{0,1\}$ and $c_{v,a} \in \{0,1\}$, defined as in Section~\ref{sec:background}. \\

\noindent
$
\begin{aligned}
& {\text{Minimize }} 
 \sum_{v,a} \alpha (1-p_{v,a}) w_{v,a} + (1-\alpha) c_{v,a}  \\
& \text{subject to } \\
& p_{v,a} + p_{u,a} - p_{p,a} \geq 0\mbox{ for $(p,u),\ (p,v) \in E(T)$ }& (c_1)\\
& p_{v,a} + p_{u,a} - p_{p,a} \leq 1\mbox{ for $(p,u),\ (p,v) \in E(T)$}& (c_2)\\
& p_{v,a} + p_{u,a}+c_{v,a} \leq 2\mbox{ for $(u,v) \in E(T)$ }& (c_3)\\
& p_{v,a} + p_{u,a}-c_{v,a} \geq 0\mbox{ for $(u,v) \in E(T)$ }& (c_4)\\
& p_{v,a} - p_{u,a}+c_{v,a} \geq 0\mbox{ for $(u,v) \in E(T)$ }& (c_5)\\
& -p_{v,a} + p_{u,a}+c_{v,a} \geq 0\mbox{ for $(u,v) \in E(T)$ }& (c_6)\\
& \sum_{a=(x_t,y)} p_{v,a} \leq 1 \text{ and } \sum_{a=(x_h,y)} p_{v,a} \leq 1\\
& \mbox{ for any marker $x$ and node $v$}& (c_7)\\
\end{aligned}
$

\medskip 
The constraints ensure parsimony ($c_1$ and $c_2$), consistency of the solution ($c_7$) and define the correct value for $c_{v,a}$ dependent on the value of $p_{a}$ along an edge $(u,v)$ ($c_{3-6}$).
This ILP has obviously a size that is polynomial in the size of the problem.

\subsection{Sampling co-optimal labelings}
The Sankoff-Rousseau DP algorithm can easily be modified to sample uniformly from the space of all optimal solutions to the Weighted SCJ labeling Problem in a forward-backward fashion.  The principle is to proceed in two stages: first, for any pair $(v,a)$ we compute the number of optimal solutions under this label for the subtree rooted at $v$. Then, when computing an optimal solution, if a DP equation has several optimal choices, one is randomly picked according to the distribution of optimal solutions induced by each choice (see Appendix for more details). This classical dynamic pro\-gram\-ming approach leads to the following result.\\

\begin{theorem}\label{thm:sampling}
  The Weighted SCJ Sampling Problem can be solved in worst-case time $O(nN(1+D)^{2M})$ and space $O(nN(1+D)^{M})$.\\
\end{theorem}

For subproblems that are too large for being handled by the Sankoff-Rousseau algorithm, the SCJ Small Parsimony Gibbs sampler recently introduced~\cite{miklos2015} can easily be modified to incorporate prior weights, although there is currently no proven property regarding its convergence.

\subsection{Weighting ancestral adjacencies}
\label{weights}
A first approach to assign weights to ancestral adjacencies consist in considering evolutionary scenarios for an adjacency independently of the other adjacencies.  An evolutionary scenario for an adjacency is a labeling of the internal nodes of the species phylogeny $T$ by the
presence or absence of the adjacency, and the parsimony score of a
scenario is the number of gains/losses of the adjacency along the branch
of $T$, \ie the SCJ score for this single adjacency. 
For a scenario $\sigma$, we denote by $p(\sigma)$ its parsimony score. Its Boltzmann score is
then defined as $\displaystyle B(\sigma)=e^{-\frac{p(\sigma)}{kT}}$, where $kT$
is a given constant. If we denote the set of all possible evolutionary scenarios for the adjacency $\{x,y\}$ by $\mathcal{S}(x,y)$, the
partition function of the adjacency and its Boltzmann probability are defined as
$$Z(x,y)=\sum_{\sigma\in \mathcal{S}(x,y)}B(\sigma),\ Pr(\sigma)=\frac{B(\sigma)}{Z(x,y)}.$$
The weight of the adjacency at internal
node~$v$ is then the sum of the Boltzmann probabilities of all
scenarios where the adjacency is present at node~$v$.
All such quantities can be computed in polynomial
time~\cite{chauve2014evolution}.

Parameter $kT$ is useful to skew the Boltzmann probability
distribution: If $kT$ tends to zero, parsimonious scenarios are
heavily favored and the Boltzmann probability distribution tends to
the uniform distribution over optimal scenarios, while when $kT$ tends
to $\infty$, the Boltzmann distribution tends toward the uniform
distribution over the whole solution space. In our experiments, we
chose a value of $kT=0.1$ that favors parsimonious scenarios but
considers also slightly suboptimal scenarios.

When aDNA sequence data is available for one or several ancestral genomes, markers identified in extant species can be related to assembled contigs of the ancestral genome, 
as in~\cite{rajaraman2013fpsac} for example. 
For an ancestral adjacency in a species for which aDNA reads are available, it is then possible to associate a sequence-based weight to the adjacency  -- either through gap filling methods (see Section~\ref{sec:results}, where we use the probabilistic model of  GAML~\cite{bovza2014gaml}), or scaffolding methods such as BESST~\cite{sahlin2014besst} for example. In comparison to the weighting approach described above, these weights are then not directly based on the underlying phylogeny, but provide an external signal for the confidence of adjacencies at the respective internal node.

%
%
\section{Results}
\label{sec:results}

We evaluated our algorithm on a simulated dataset and compared its sensitivity and precision to several other reconstruction methods.
Further, we applied our method to two real datasets: \textit{mammalian} and \textit{Yersinia} genomes. The mammalian dataset was used in the studies~\cite{chauve2008methodological} and~\cite{miklos2015}. It contains six mammalian species and two outgroups, spanning over $100$ million years of evolution, and five different marker sets of varying resolution (minimal marker length). 
Our experimental results consider issues related to the complexity of our algorithm, 
the use of a pure SCJ reconstruction (obtained when the $\alpha$ parameter equals $0$)
and the relative impact of the value of $\alpha$ on both the total evolutionary cost and the ancestral gene orders fragmentation. 
Our second dataset contains eleven \textit{Yersinia} genomes, an important human pathogen. 
This dataset contains contigs from the recently sequenced extinct agent of the Black Death pandemic~\cite{bos2011draft} that occurred roughly $650$ years ago.
We refer to Supplementary Fig.~\ref{fig:trees1} and~\ref{fig:trees2} for the species phylogenies of these two datasets.

\subsection{Simulations}
We created simulated datasets as described in~\cite{feijao2015reconstruction}: with a birth-rate of $0.001$ and a death rate of $0$, we simulated $20$ binary trees with $6$ leaves and scaled the branch lengths such that the tree has a diameter $D=2n$, where $n$ is the number of markers in each unichromosomal genome. The root genome with $n = 500$ markers is then evolved along the branches of the tree by applying inversions and translocations with a probability of $0.9$ and $0.1$ respectively. The number of rearrangements at each branch corresponds to the simulated branch length, the total number of rearrangements ranges from $1242$ to $2296$ in the simulated trees.
We compare results of our implementation \textit{PHYSCA} for different values of $\alpha \in \{0,0.3,0.5,0.8,1\}$ with the tools \textit{RINGO}~\cite{feijao2015reconstruction}, \textit{MGRA}~\cite{avdeyev2016reconstruction}, Fitch-SCJ~\cite{biller2013rearrangement}, \textit{ROCOCO}~\cite{STO-WIT-2009, WIT-2010} (dense approach for signed adjacencies) and \textit{ANGES}~\cite{jones2012anges} (adjacencies only). We computed adjacency weights as described in~\autoref{weights} with the software \textit{DeClone}~\cite{chauve2014evolution} and parameter $kT \in \{0.1,1\}$.

The methods \textit{RINGO} and \textit{MGRA} are global approaches minimizing the DCJ-distance in the tree, while \textit{ANGES} reconstructs specific ancestors locally in the tree and is applied for each node separately. For $\alpha = 0$, our objective is finding a consistent, most parsimonious solution and equals the objectives of Fitch-SCJ and \textit{ROCOCO},  where Fitch-SCJ always finds the most fragmented solution whereas \textit{ROCOCO} and our method aim at reporting least fragmented reconstructions.

\begin{figure}[t]
\begin{minipage}[t]{0.49\linewidth}
\includegraphics[width=\linewidth]{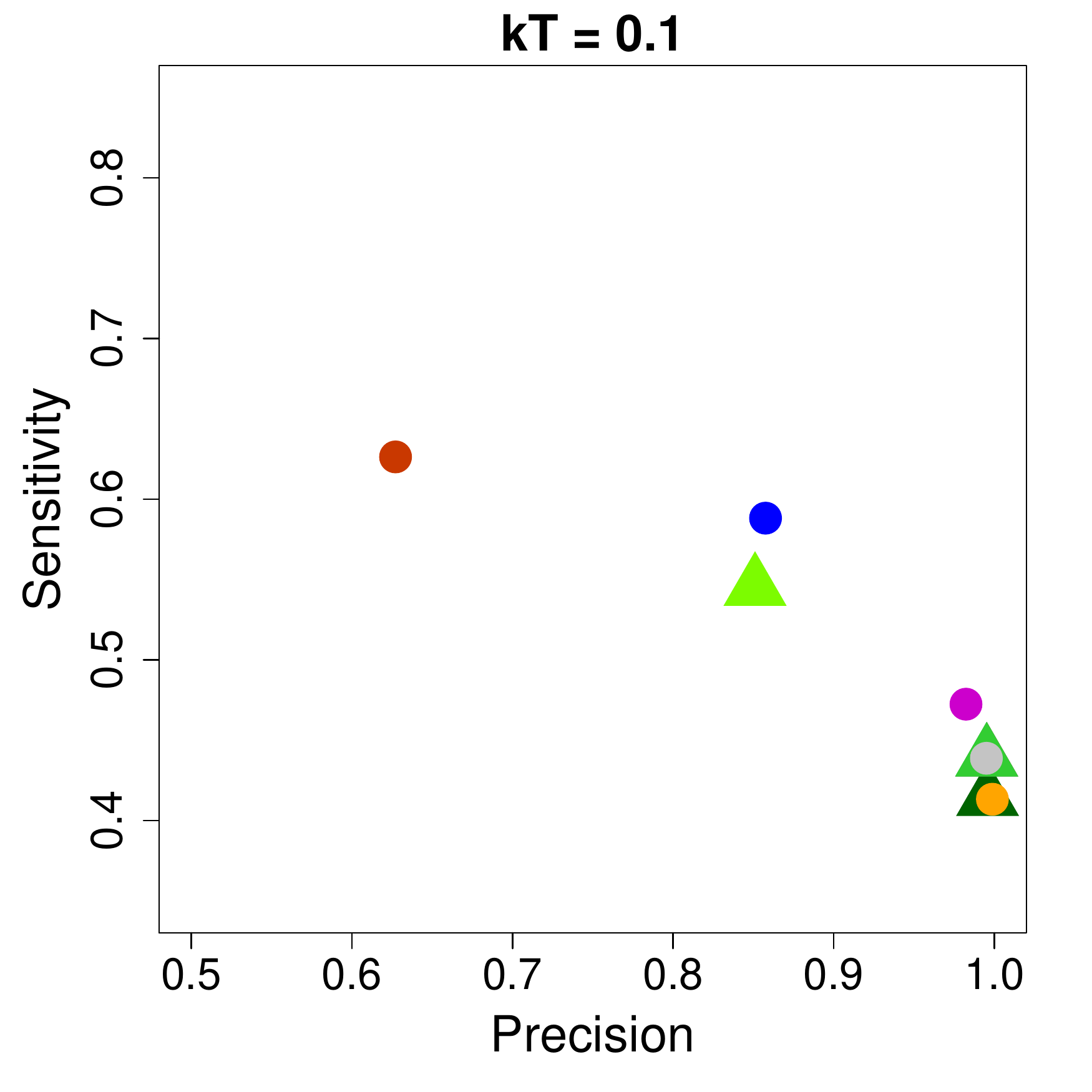}
\end{minipage}
\hfill
\begin{minipage}[t]{0.49\linewidth}
\includegraphics[width=\linewidth]{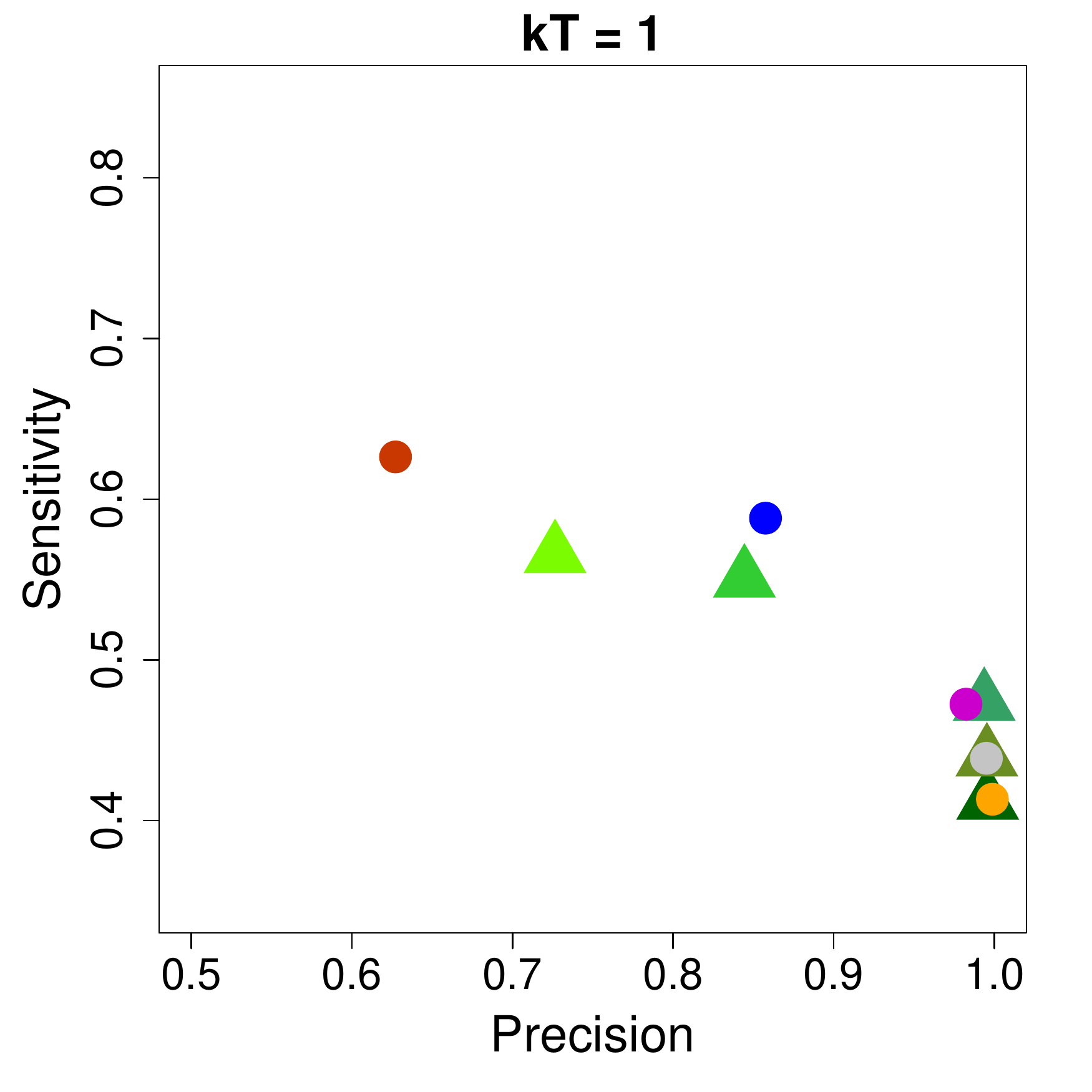}
\end{minipage}
\hfill
\begin{minipage}[t]{0.49\linewidth}
\includegraphics[width=\linewidth]{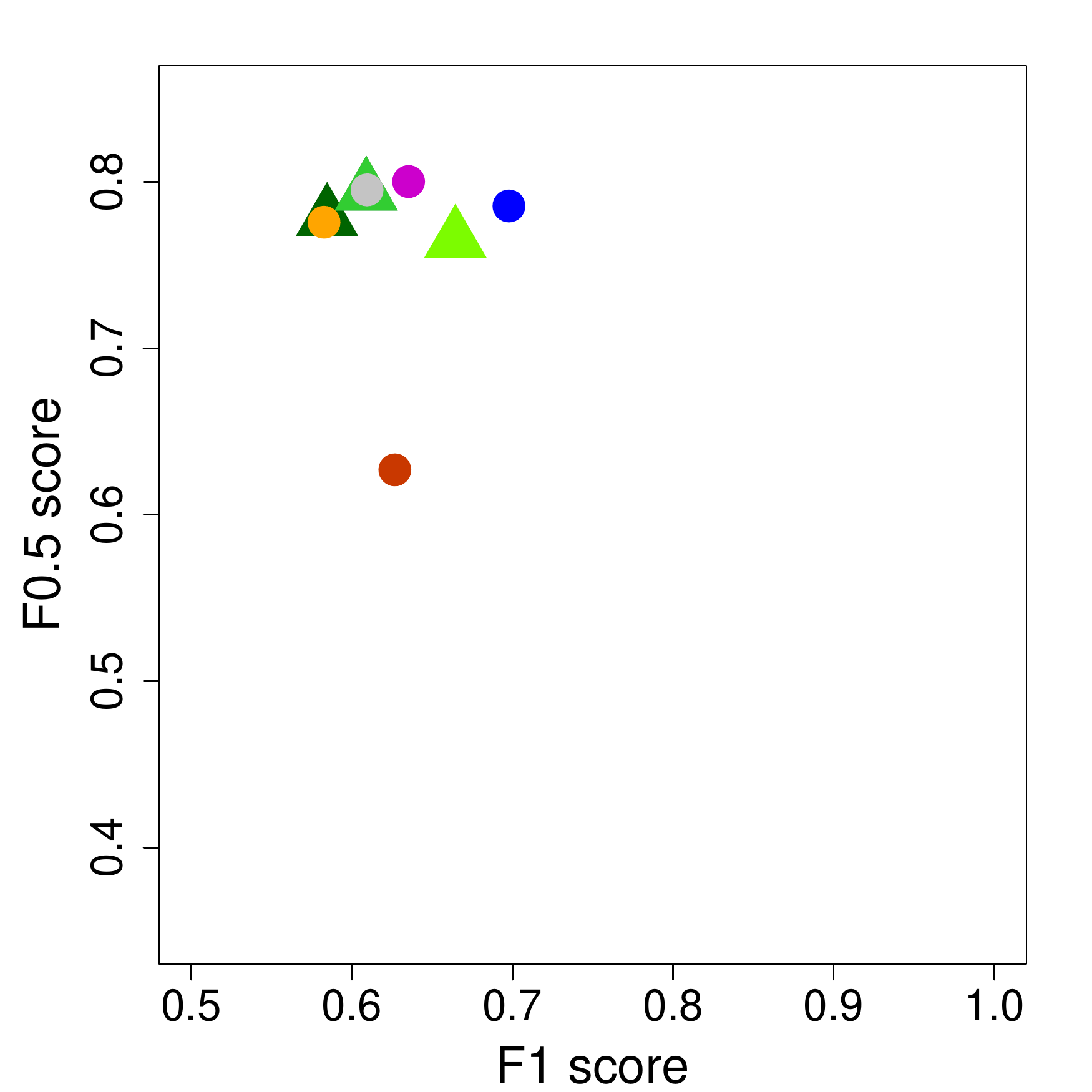}
\end{minipage}
\hfill
\begin{minipage}[t]{0.49\linewidth}
\includegraphics[width=\linewidth]{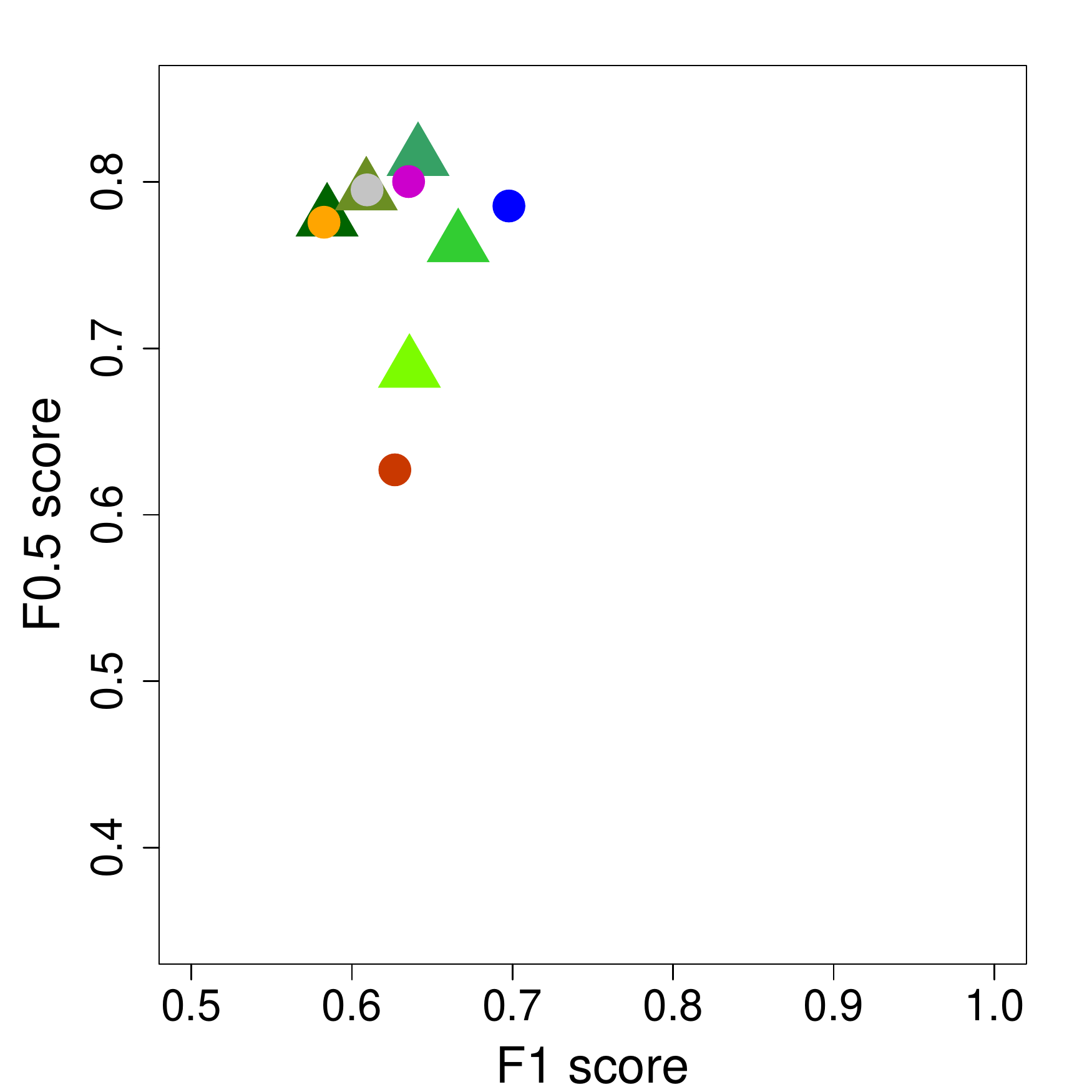}
\end{minipage}
\begin{minipage}[t]{\linewidth}
\vspace{0.0000001cm}
\centering
\includegraphics[width=0.6\linewidth]{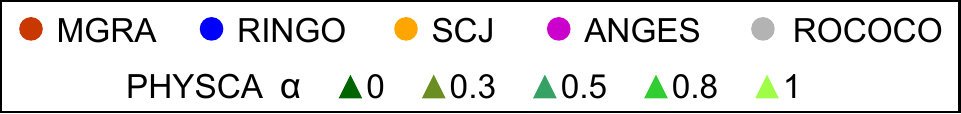}
\end{minipage}
\caption{Average precision and sensitivity (top), and $F_1$ and $F_{0.5}$ (bottom) of reconstructions on $20$ simulated datasets. Adjacency weights have been obtained with parameters $kT = 0.1$ (left) and $kT = 1$ (right). 
}
\label{fig:simulation}
\end{figure}

We measured sensitivity and precision of the reconstructions based on the comparison of simulated and reconstructed adjacencies by the different methods.
A high sensitivity indicates the ability to recover the true marker order of ancestors in the phylogeny, while a high precision denotes few wrongly reconstructed adjacencies. 
As shown in~\autoref{fig:simulation}, our method reaches a high precision of $0.99$ for all values of $\alpha \geq 0.5$, while increasing the sensitivity in comparison to the pure Fitch-SCJ solution by reducing the fragmentation of the reconstructed scaffolds. For higher values of $\alpha$, the influence of the weighting becomes apparent: for $kT = 0.1$, the precision only decreases for $\alpha = 1$, while for $kT = 1$, the precision decreases also for 
lower values of $\alpha$, however leading to more complete reconstructions. 
In comparison, both DCJ-based methods \textit{RINGO} and \textit{MGRA} produce less fragmented solutions by recovering more true adjacencies under the jeopardy of also reconstructing more false adjacencies. The sensitivity and precision of Fitch-SCJ, \textit{ROCOCO} and \textit{ANGES} are comparable to our method for low to medium values of $\alpha$.

The $F_{1}$ score assesses the relation of sensitivity and precision with equal importance. \textit{RINGO} achieves a better $F_{1}$ score than all other methods. 
The $F_{0.5}$ score emphasizes the precision of a method over its sensitivity. With this measure, our method with $kT = 1$ and $\alpha = 0.5$ outperforms the other tools, while \textit{ROCOCO} and \textit{ANGES} also reach similarly good scores.

In general, it can be seen that the equal contribution of global evolution and local adjacency weights in the objective function provides a reliable reconstruction and further a useful tool to explore the solution space under different values of $\alpha$.

\subsection{Mammalian dataset}
We used the markers computed in~\cite{chauve2008methodological} from
whole-genome alignments. The extant species contain a diverse number of chromosomes ranging from 9 chromosomes in \textit{opossum} to 39 chromosomes in \textit{pig}.  
Unique and universal markers were computed as synteny blocks with different resolution in terms of minumum marker length. Note that all rearrangement breakpoints are therefore located outside of marker coordinates.
It results in five different datasets
varying from $2,185$ markers for a resolution of $100\,$kb to
$629$ markers for a resolution of $500\,$kb.

We considered all adjacencies present in at least one extant genome as potentially ancestral. To weight an adjacency at all internal nodes of the tree, we relied on evolutionary scenarios for each single adjacency, in terms of gain/loss, independently of the other adjacencies (\ie without considering consistency of ancestral marker orders). We obtain these weights using the software {DeClone}~\cite{chauve2014evolution}, and we refer to them as \textit{DeClone weights}. 
We considered two values of the DeClone parameter $kT$, $0.1$ and $1$, the former ensuring that only adjacencies appearing in at least one optimal adjacency scenario have a significant DeClone weight, while the latter samples adjacencies outside of optimal scenarios.
For the analysis of the ancestral marker orders obtained with our algorithm, we considered the data set at $500$\,kb resolution and sampled $500$ ancestral marker orders for all ancestral species under different values of $\alpha$.  

\subsubsection*{Complexity}
\begin{figure}
\centering
\includegraphics[width=0.9\linewidth]{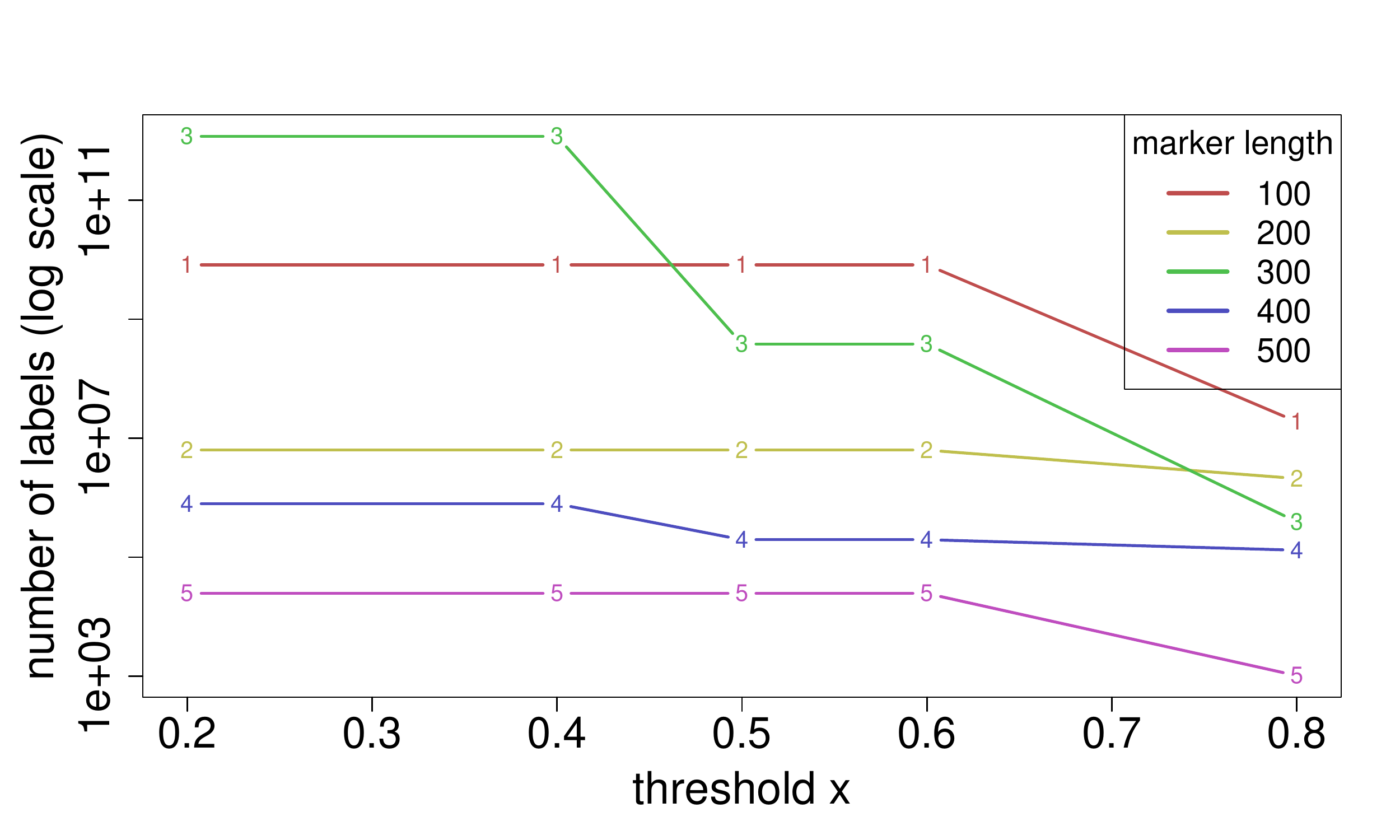}
\caption{Number of different labels for the largest connected component in each of the mammalian datasets. This statistic provides an upper bound for the actual complexity of our reconstruction algorithm.}
\label{numberLabels}
\end{figure}

The complexity of our algorithm is dependent on the size of the largest connected component of the global adjacency graph. In order to restrict the complexity, we kept only adjacencies whose weights are above a given threshold~$x$.
Figure~\ref{numberLabels} shows the expected decrease in computational complexity correlated to threshold~$x$ for the five different minimal marker lengths. 
In most cases, all connected components are small enough to be handled by our exact algorithm in reasonable time except for very large components in the marker sets with higher resolution under a low threshold $x$. For the $500$\,kb dataset with $x=0.2$ and $kT=1$, the computation of one solution takes on average 200\,s on a 2.6\,GHz i5 with 8\,GB of RAM. It can be reduced to 30\,s when DeClone weights are based on $kT=0.1$. 
This illustrates that our algorithm, despite an exponential worst-case time complexity, can process realistic datasets in practice.
%

\begin{figure}
\centering
\includegraphics[width=0.9\linewidth]{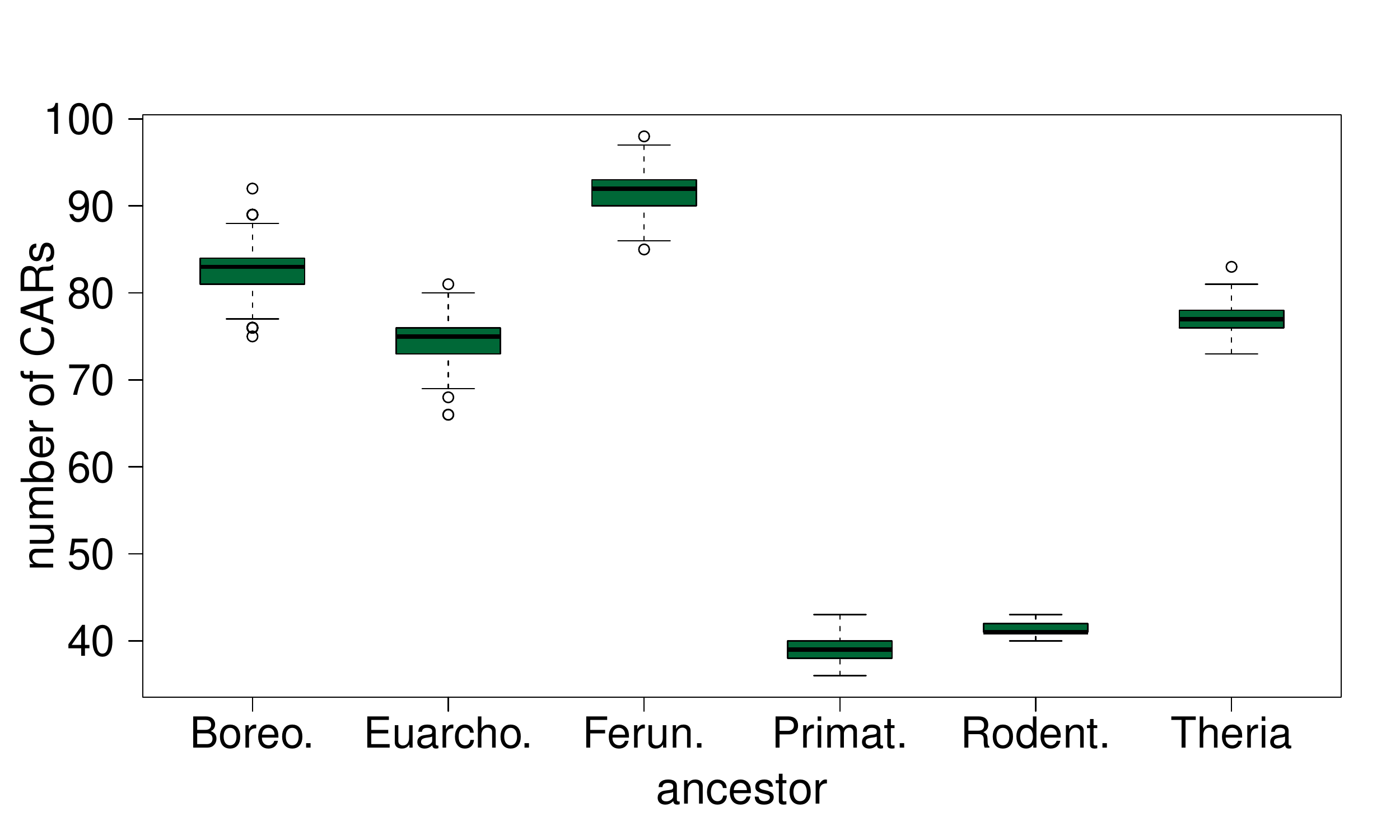}
\caption{Number of reconstructed CARs at each internal node in 500 samples for the mammalian dataset with $500$\,kb resolution, $x = 0.2$ and $\alpha = 0$. }
\label{fig:fragmentationAlpha0}
\end{figure}

\subsubsection*{Optimal SCJ labelings}

Next, we analyzed the $500$ optimal SCJ labelings obtained for $\alpha=0$, \ie aiming only at minimizing the SCJ distance, and considered the fragmentation of the ancestral gene orders (number of CARs) and the total evolutionary distance. Note that, unlike the Fitch algorithm used in~\cite{feijao2011scj}, our algorithm does not favor fragmented assemblies by design but rather considers all optimal labelings. Sampling of co-optimal solutions shows that the pure SCJ criterion leads to some significant variation in terms of number of CARs (Figure~\ref{fig:fragmentationAlpha0}).  
In contrast,  Table~\ref{tab:robustnessAlpha0} shows that most observed ancestral adjacencies are present in all sampled scenarios. About $5\%$ of adjacencies, mostly located at nodes higher up in the phylogeny, are only present in a fraction of all sampled scenarios, indicating that there is a small number of conflicts between potential adjacencies that can be solved ambiguously at the same parsimony cost.

The optimal SCJ distance in the tree for $\alpha=0$ is $1,674$, while the related DCJ distance in the sampled reconstructions varies between $873$ and $904$ (Figure~\ref{fig:SCJAllAlpha}). In comparison, we obtained a DCJ distance of $829$ with GASTS~\cite{xu2011gasts}, a small parsimony solver directly aiming at minimizing the DCJ distance. More precisely, over all ancestral nodes, $70$ adjacencies found by GASTS do not belong to our predefined set of potential ancestral adjacencies and another $147$ appear in the $500$ samples with a frequency below $50\%$. 
This illustrates both a lack of robustness of the pure SCJ optimal labelings, and some significant difference between the SCJ and DCJ distances.

\begin{table}
\centering
\caption{Frequency of adjacencies in 500 samples with $\alpha = 0$ as percentage of optimal labelings they appear in.}
\begin{tabular}{lccc}\toprule
Ancestor & \multicolumn{3}{c}{Frequency $f$}\\
 & $f=100\%$ & $100\%>f>50\%$ & $f<50\%$ \\
\midrule
\it Boreoeutheria & 94.66 & 1.07 & 4.27\\
\it Euarchontoglires & 95.42 & 0.88 & 3.79\\
\it Ferungulates & 96.53 & 0.55 & 2.92\\
\it Primates & 98.82 & 0.34 & 0.84\\
\it Rodentia & 99.49 & 0.34 & 0.17\\
\it Theria & 97.67 & 0.89 & 1.43\\
root node & 92.23 & 1.23 & 6.53\\
\bottomrule
\end{tabular}
\label{tab:robustnessAlpha0}
\end{table}

Finally, we compared the Boltzmann probabilities of ancestral adjacencies (DeClone weights) with the frequency observed in the $500$ samples. There is a very strong agreement for DeClone weights obtained with $kT=0.1$ as only $14$ ancestral adjacency have a DeClone weight that differs more than $10\%$ from the observed frequency in the samples. This shows that, despite the fact that the DeClone approach disregards the notion of conflict, it provides a good approximation of the optimal solutions of the SCJ Small Parsimony Problem.

\subsubsection*{Ancestral reconstruction with DeClone weights and varying values of $\alpha$.}
\begin{figure}
\centering
\includegraphics[width=0.9\linewidth]{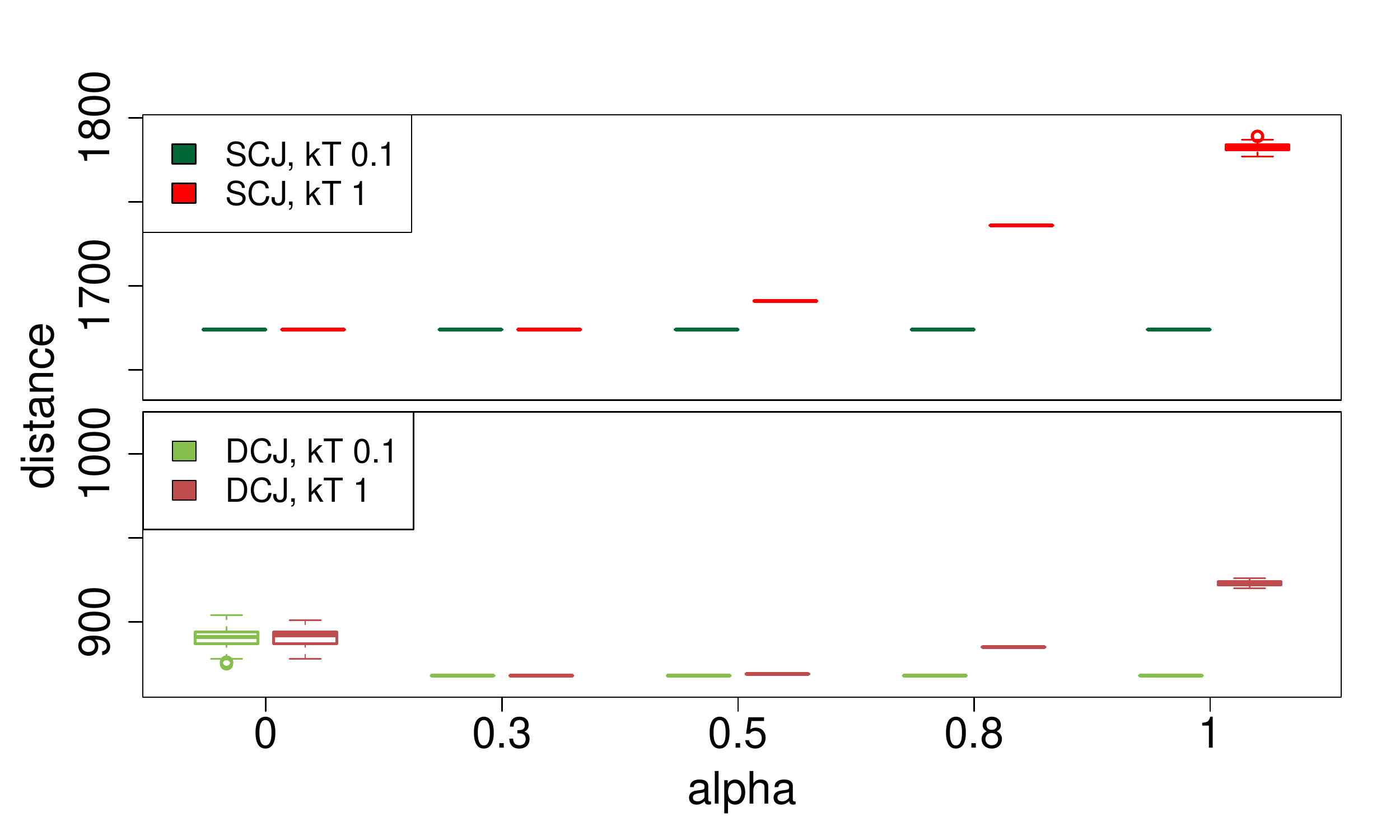}
\caption{SCJ distance (upper half) and DCJ (lower half) distance in the whole tree for all samples and selected values of $\alpha$ in the mammalian dataset.}
\label{fig:SCJAllAlpha}
\end{figure}

\begin{figure}
\centering
\includegraphics[width=0.9\linewidth]{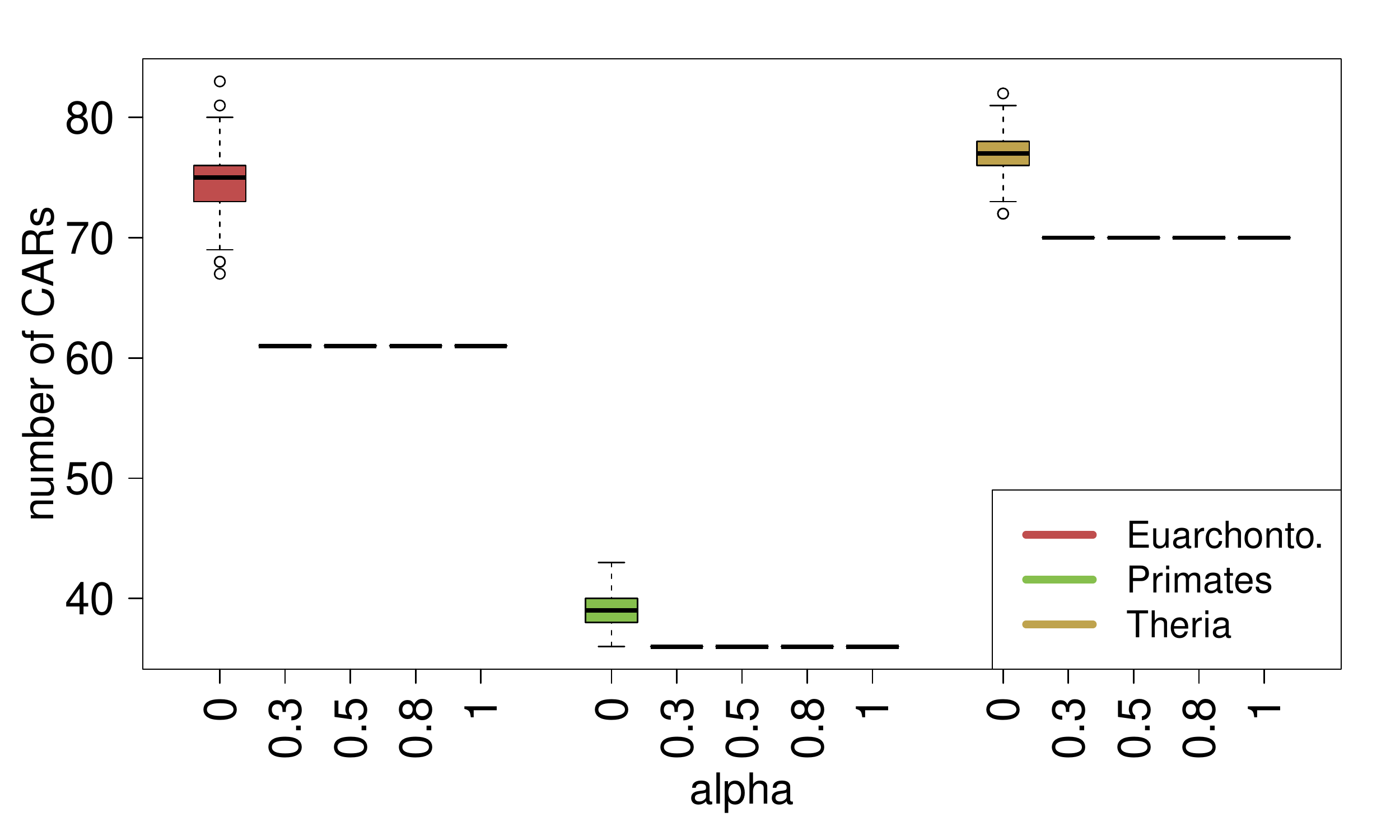}
\caption{Number of CARs in the mammalian dataset in all samples at selected internal nodes for different values of $\alpha$ reconstructed with DeClone weights under $kT=0.1$. While the number of CARs differs in the case of $\alpha = 0$ where the adjacency weights are not considered, the fragmentation stays constant for the other values of $\alpha$.}
\label{fig:fragmentation}
\end{figure}

For $\alpha > 0$, our method minimizes a combination of the SCJ distance with the DeClone weights of the adjacencies discarded to ensure valid ancestral gene orders.  Again, we sampled $500$ solutions each for different values of $\alpha$ with the $500$\,kb data set. 
We distinguish between DeClone parameter $kT=0.1$ and $kT=1$. Figures~\ref{fig:SCJAllAlpha} and~\ref{fig:fragmentation} show the respective observed results in terms of evolutionary distance and fragmentation.
%
%

For $kT=0.1$, the optimal SCJ and DCJ distance over the whole tree hardly depends on $\alpha$.  
Including the DeClone weights in the objective actually results in the same solution, independent of $\alpha>0$. In fact, while applying a low weight threshold of $x=0.2$, the set of potential adjacencies is already consistent at all internal nodes except for a few conflicts at the root that are solved unambiguously for all values of $\alpha$. This indicates that building DeClone weights on the basis of mostly optimal adjacency scenarios (low $kT$) results in a weighting scheme that agrees with the evolution along the tree for this dataset. More importantly, Figures~\ref{fig:SCJAllAlpha} and ~\ref{fig:fragmentation} show that the combination of DeClone weights 
followed by our algorithm, leads to a robust set of ancestral gene orders.

In comparison, for $kT=1$, we see an increase in SCJ and DCJ distance for higher $\alpha$, while the number of CARs at internal nodes decreases, together with a loss of the robustness of the sampled optimal results when $\alpha$ gets close to $1$. It can be explained by the observation that the weight distribution of ancestral adjacencies obtained with DeClone and $kT=1$ is more balanced than with $kT=0.1$ as it considers suboptimal scenarios of adjacencies with a higher probability. 
It further illustrates that, when the global evolutionary cost of a solution has less weight in the objective function, the algorithm favors the inclusion of an adjacency of moderate weight that joins two CARs while implying a moderate number of evolutionary events (for example an adjacency shared by only a subset of extant genomes). 
From that point of view, our algorithm -- being efficient enough to be run on several values of $\alpha$ -- provides a useful tool to evaluate the relation between global evolution and prior confidence for adjacencies whose pattern of presence/absence in extant genomes is mixed.

\subsection{\textit{Yersinia pestis} dataset}
We started from fully assembled DNA sequences of seven \textit{Yersinia pestis} and four \textit{Yersinia pseudotuberculosis} genomes. In addition, we included aDNA single-end reads and $2\,134$ contigs of length~$>500$bp assembled from these reads for the Black Death agent, considered as ancestral to several extant strains~\cite{bos2011draft}. We refer to this augmented ancestral node as the \emph{Black Death (BD) node}.
The marker sequences for all extant genomes were computed as described in~\cite{rajaraman2013fpsac},  restricting the set of markers to be unique and universal. 
We obtained a total of $2,207$ markers in all extant genomes and $2,232$ different extant adjacencies, thus showing a relatively low level of syntenic conflict compared to the number of markers, although it implies a highly dynamic rearrangement history over the short period of evolution~\cite{rajaraman2013fpsac}.

As for the mammalian dataset, we considered as potentially ancestral any adjacency that appears in at least one extant genome. However for this dataset, reducing the complexity by applying a weight threshold~$x$ was not necessary.
For the BD node, adjacency weights can be based on the given aDNA reads for a given potential ancestral adjacency as follows. First, we used FPSAC~\cite{rajaraman2013fpsac} to compute DNA sequences filling the gaps between any two adjacent marker extremities (obtained by aligning the gap sequences of the corresponding conserved extant adjacencies and reconstructing a consensus ancestral sequence using the Fitch algorithm). 
Then we computed the weights as a likelihood of this putative gap sequence given the aDNA reads, using the GAML probabilistic model described in~\cite{bovza2014gaml}. Each adjacency together with its template gap sequence details a proposition for an assembly~$A$ as a piece of the real ancestral sequence, and given the aDNA read set~$R$, the model then defines a probability $Pr(R|A) = \prod_{r \in R} Pr(r|A)$ for observing the reads~$R$ given that $A$ is the correct assembly. The probability $Pr(r|A)$ can be computed by aligning $r$ to the assembly~$A$ while the alignment is evaluated under an appropriate sequencing error model. We refer to~\cite{bovza2014gaml} for details. 

\subsubsection*{Ancestral reconstruction with aDNA weights}
Again we sampled $500$ solutions for this dataset. We computed the weights at the BD node based on the aDNA data, while adjacencies at all other nodes were given weight  $0$. Hence we can investigate the influence of including the aDNA sequencing data in the reconstruction while for the rest of the tree, the weights do not impact the objective function. Moreover, this weighting scheme addresses the issue of potential BD adjacencies with a low weight due to the difficulty of sequencing ancient DNA. 

\begin{figure}
\centering
\includegraphics[width=0.9\linewidth]{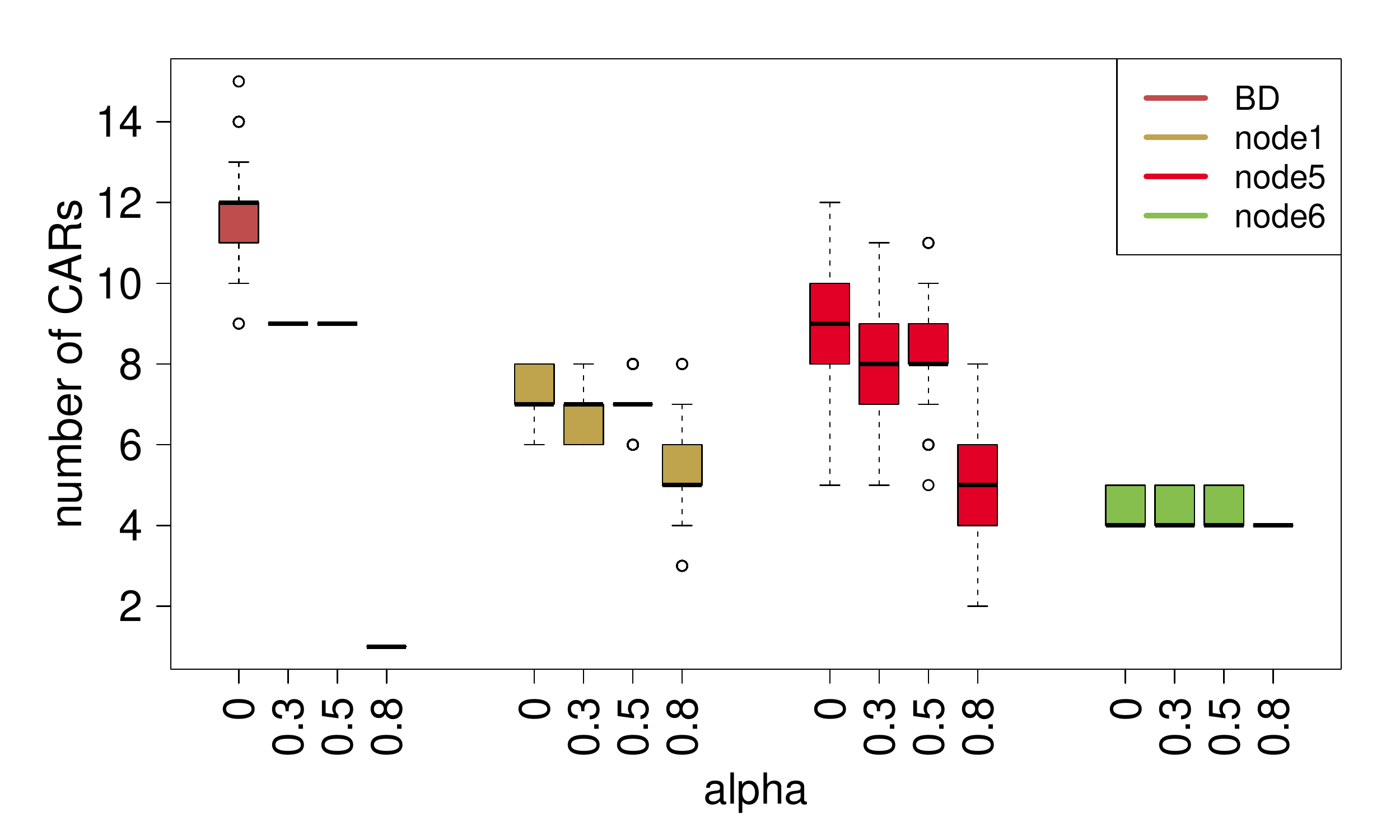}
\caption{Reconstructed number of CARs in the \textit{yersinia} dataset with aDNA weights at the BD node and 0 otherwise, for four ancestral nodes. }
\label{fig:GAML0}
\end{figure}

As shown in Figure~\ref{fig:GAML0}, for selected internal nodes of the phylogeny, the pure SCJ solutions at $\alpha = 0$ result in the highest fragmentation, while the number of CARs decreases as we increase the importance of the adjacency weights in the objective of our method. For the BD node, when including the aDNA weights, the fragmentation is decreasing while the reconstructions for each $\alpha>0$ are robust. At the other nodes, the applied sequencing weights also reduce the fragmentation except for node6 which is located in the pseudotuberculosis subtree and hence more distant to the BD node. This shows that the aDNA weights not only influence the reconstructed adjacencies at the BD node, but  also other nodes of the tree.

\section{Conclusion}
\label{sec:conclusion}

Our main contributions are the introduction of the Small Parsimony
Problem under the SCJ model with adjacency weights, together
with an exact parameterized algorithm for the optimization and
sampling versions of the problem. The motivation for this
problem is twofold: incorporating sequence signal from aDNA data when it is available, and recent works showing that the reconstruction of ancestral genomes through the independent analysis of adjacencies is
an interesting approach~\cite{berard2012evolution,chauve2014evolution,feijao2011scj,miklos2015}. 

Regarding the latter motivation, we address a general issue of these approaches that either
ancestral gene orders are not consistent or are quite fragmented if the methods are constrained to ensure consistency. The main idea
we introduce is to take advantage of sampling approaches recently
introduced in~\cite{chauve2014evolution}  to weight potential ancestral adjacencies and thus
direct, through an appropriate objective function, the reconstruction
of ancestral gene orders.  Our results on the mammalian dataset suggest that this approach leads to a robust ancestral genome structure. However, we can observe a significant difference with a DCJ-based ancestral reconstruction, a phenomenon that deserves to be explored further.
Our algorithm, which is based on the Sankoff-Rousseau
algorithm similarly to several recent ancestral reconstruction
algorithms~\cite{berard2012evolution,chauve2014evolution,miklos2015}, is a
parameterized algorithm that can handle real instances containing a
moderate level of syntenic conflict. 
Our experimental results on both
the mammalian and bacterial datasets suggest that introducing prior
weights on adjacencies in the objective function has a significant
impact in reducing the fragmentation of ancestral gene orders, even
with an objective function with balanced contributions of the SCJ
evolution and adjacency weights. 
For highly conflicting instances, it can be discussed if a reconstruction through small parsimony is the right
approach to solve these conflicts or if these should be addressed
differently.  

Our sampling algorithm improves on the
Gibbs sampler introduced in~\cite{miklos2015} in terms of
computational complexity and provides a useful tool to study ancestral
genome reconstruction from a Bayesian perspective. Moreover, our algorithm is flexible regarding the potential ancestral gene adjacencies provided as input and could easily be associated with other ideas, such as intermediate genomes for example~\cite{feijao2015reconstruction}. 

There are several research avenues opened by our work. From a theoretical point of view, we know the problem  we
introduced is tractable for $\alpha=0$ and
$\alpha=1$, and we show it is hard for $\alpha > 33/34$,
but it remains to see whether it is hard otherwise.
Further, given that the considered objective is a 
combination of two objectives to be optimized simultaneously, 
Pareto optimization is an interesting aspect that should be considered.
 Our model could also be extended towards
 other syntenic characters than adjacencies, \ie groups of more than two
 markers, following the ancient gene clusters reconstruction approach
 introduced in~\cite{STO-WIT-2009}. As ancestral gene orders are defined by consistent sets of adjacencies, the principle of our dynamic programming algorithm could be conserved and it would only be a matter of integrating gene clusters into the objective function. 
From a more applied point of view, one would like to
incorporate duplicated and deleted markers into our Small Parsimony
Problem. There exist efficient algorithms for the case of a single
adjacency~\cite{berard2012evolution,chauve2014evolution} that can provide
adjacency weights, and natural extensions of the SCJ model to
incorporate duplicated genes. 
However it remains to effectively combine these ideas.  
 Finally, again due to the flexibility and simplicity of the Sankoff-Rousseau dynamic programming algorithm, 
 one could easily extend our method towards the inference of extant adjacencies if some extant genomes are provided in partially assembled form following the general approach described in
 ~\cite{aganezov2015scaffold,anselmetti2015}. 
 
 This would pave the way towards a fully integrated phylogenetic scaffolding method that combines evolution and sequencing data for selected extant and ancestral genomes.

\subsection*{Acknowledgements}
NL and RW are funded by the International DFG Research Training Group GRK 1906/1. CC is funded by NSERC grant RGPIN-249834.
%
\bibliography{references_shortened}{}
\bibliographystyle{splncs03}
\clearpage
\appendix
 \label{appendix}
 \subsection*{Methods: supplementary material}
\begin{figure}[h]
\centering
  \includegraphics[width=0.65\linewidth]{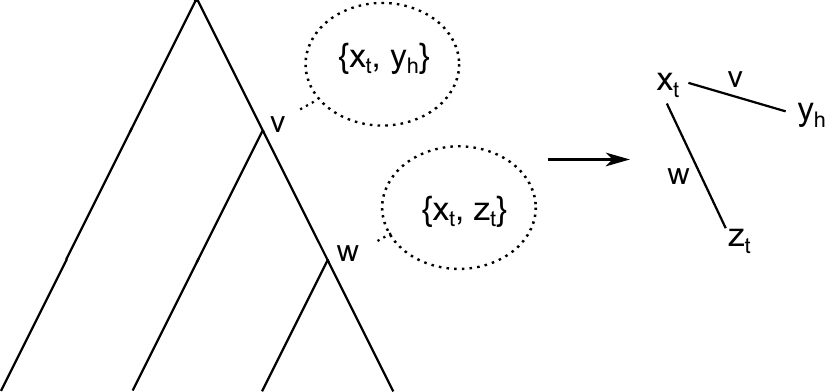}
  \caption{Simplified example of a connected component in the global adjacency graph. All internal nodes of the tree are
    augmented with adjacency graphs. At node $v$, the graph contains
    adjacency $\{x_t,y_h\}$, at node $u$, the graph contains adjacency
    $\{x_t,z_t\}$. Along the edge $\{u,v\}$, one adjacency has been cut,
    the other has been joined. In the global adjacency graph, we see a
    connected component that contains two edges incident to $x_t$. This component does not
    contain a conflict, just a rearrangement. See Fig.~\ref{fig:example} below
    for a $C$ that contains a conflict.}\label{fig:gag}
\end{figure}
\begin{figure}[h]
\centering
\includegraphics[width=0.65\linewidth]{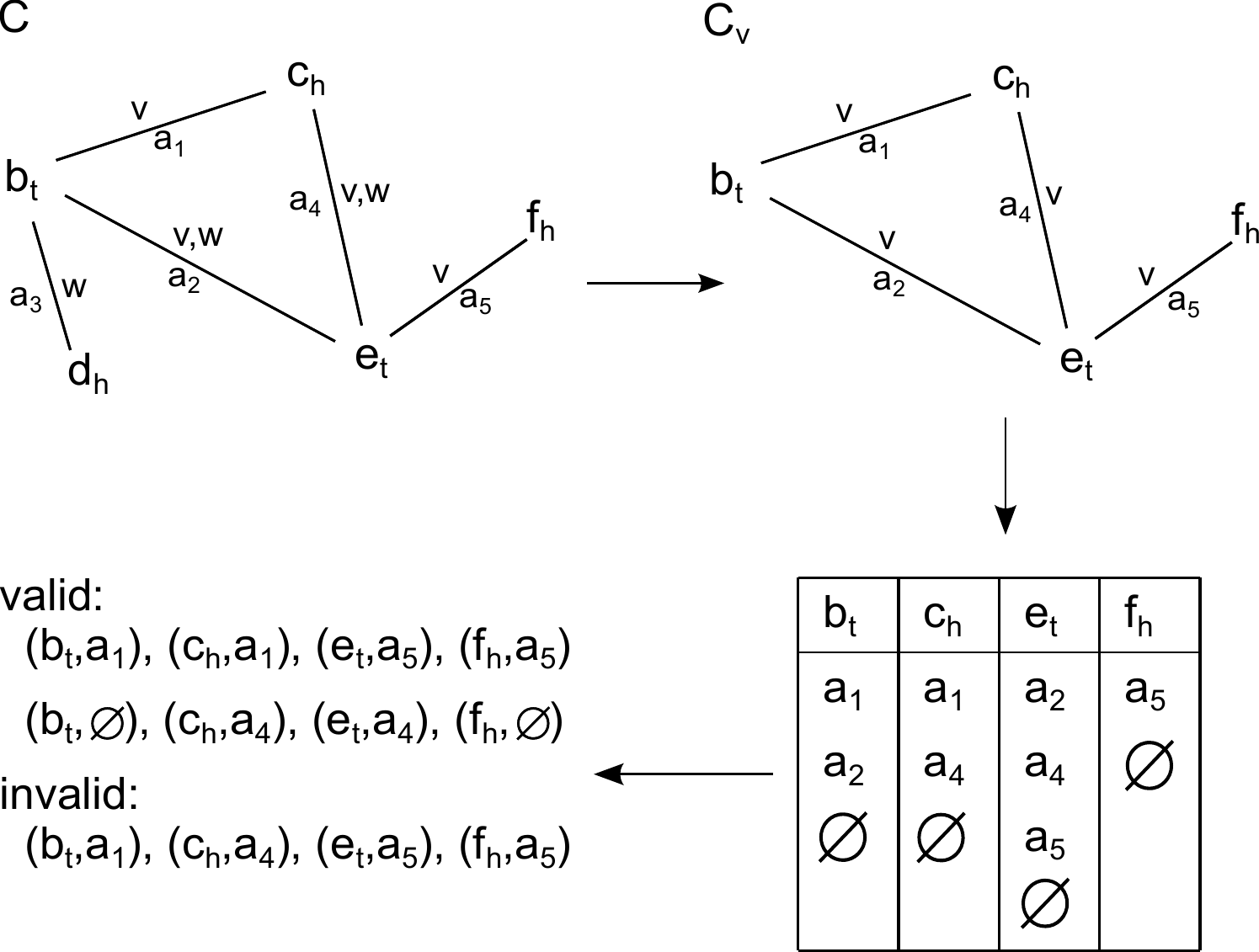}
\caption{Example for a given connected component $C$. At each node $v$ of the tree, possible assignments are defined by the connected component containing all edges annotated with $v$. Possible assignments for a marker extremity, \eg $b_t$, are defined by the incident adjacency edges, hence in this example we can \eg assign $a_1$,$a_2$ or $\emptyset$ to $b_t$. Here two valid joint labeling for $v$ are shown, while the third one assigns different adjacencies for $b_t$ and $c_h$ and is therefore invalid.}
\label{fig:example}
\end{figure}
\subsubsection*{A DP sampling algorithm.}
In the bottom-up traversal, in addition to the minimal cost induced by labeling a node~$v$ with a specific label $a \in L$, we can also store the number of optimal solutions under this label for the subtree rooted at $v$. Let $x$ and $y$ be the children of $v$, and $L_x$ and $L_y$ the sets of labels that induced the minimum value for $a$ at $v$ (which means a label out of these sets is assigned in the backtracking phase if $v$ is labeled with $a$). Then
$$\overline{C}(v,a)=\left(\sum_{l \in L_x}\overline{C}(x,l)\right)\left(\sum_{l \in L_y}\overline{C}(y,l)\right)$$
gives the number of optimal solutions for the subproblem rooted at $v$. 
At the root, we might have the choice between different labels with minimum cost.
Let $L_{root}$ be the set of these labels, then the number of overall possible co-optimal solutions is simply the sum of the number of solutions for all optimal root labels:
$\overline{C}(root)= \sum_{l \in L_{root}} \overline{C}(root,l).$
Subsequently in the top-down traversal, choose a label $l \in L_{root}$ with probability $\frac{\overline{C}(root,l)}{\overline{C}(root)}$. If at an internal node more than one label in a child node induced the minimum value, choose one of these labels analogously.
\subsection*{Results: supplementary material}
%
%
\begin{figure}[h]
\centering
\includegraphics[width=0.65\linewidth]{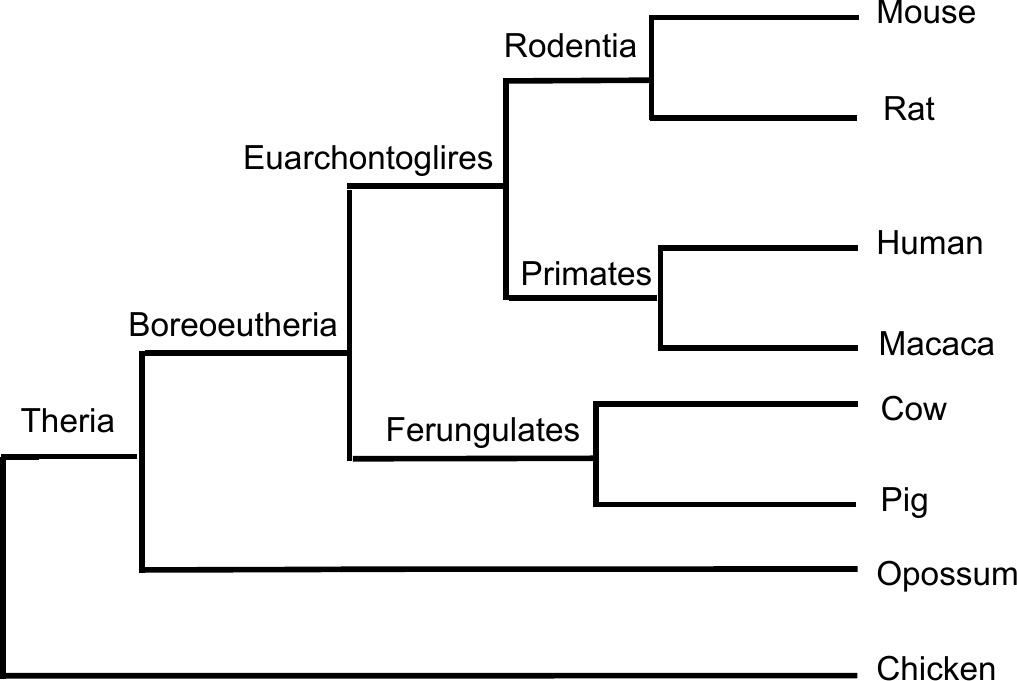}
\caption{Underlying phylogeny of the considered genomes for the mammalian dataset used in the evaluation.}\label{fig:trees1}
\vspace{0.5cm}
\includegraphics[width=0.75\linewidth]{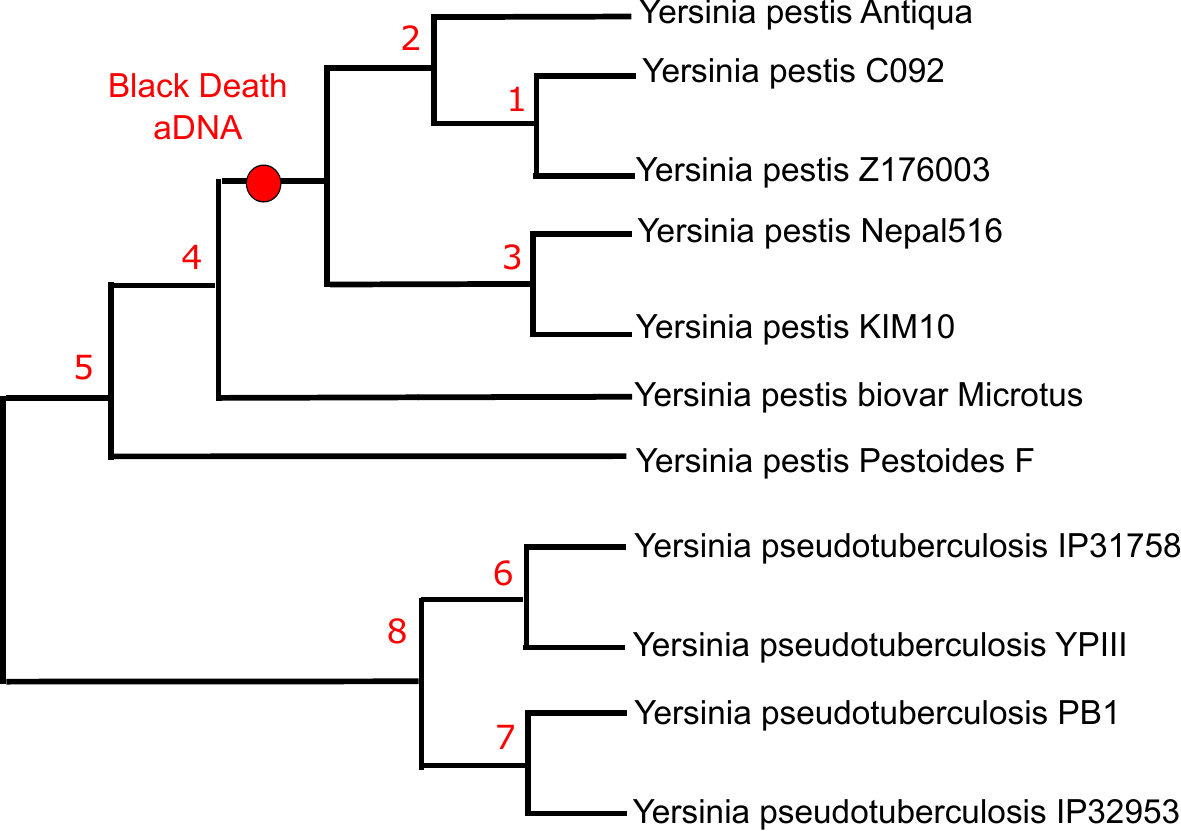}
\caption{Underlying phylogeny of the considered genomes for the \textit{Yersinia} dataset used in the evaluation.}\label{fig:trees2}
\vspace{0.3cm}
\includegraphics[width=0.73\linewidth]{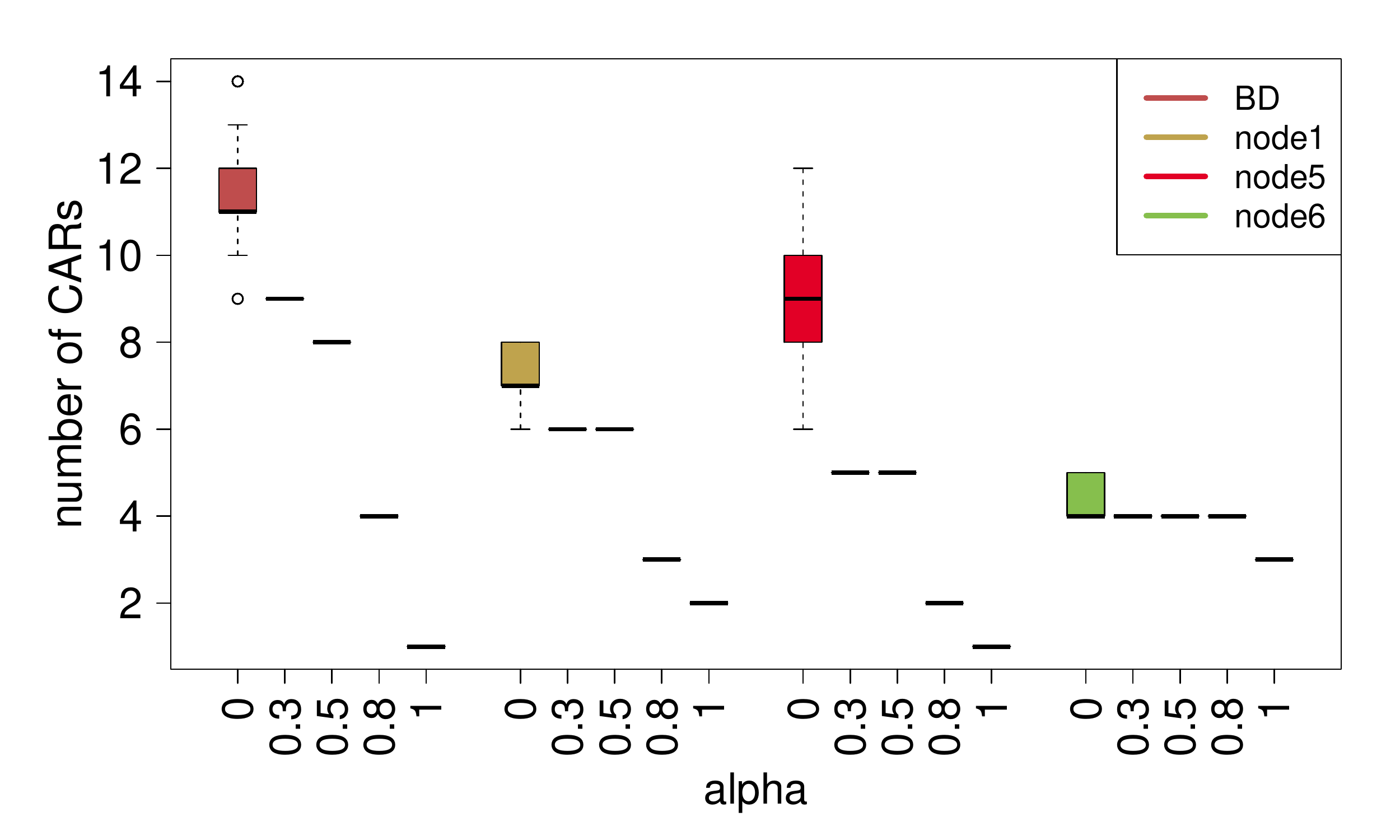}
\caption{\textit{Yersinia} dataset: Reconstructed number of CARs with DeClone weights at $kT=1$.}
\label{fig:DeClone_frag}
\vspace{0.3cm}
\includegraphics[width=0.73\linewidth]{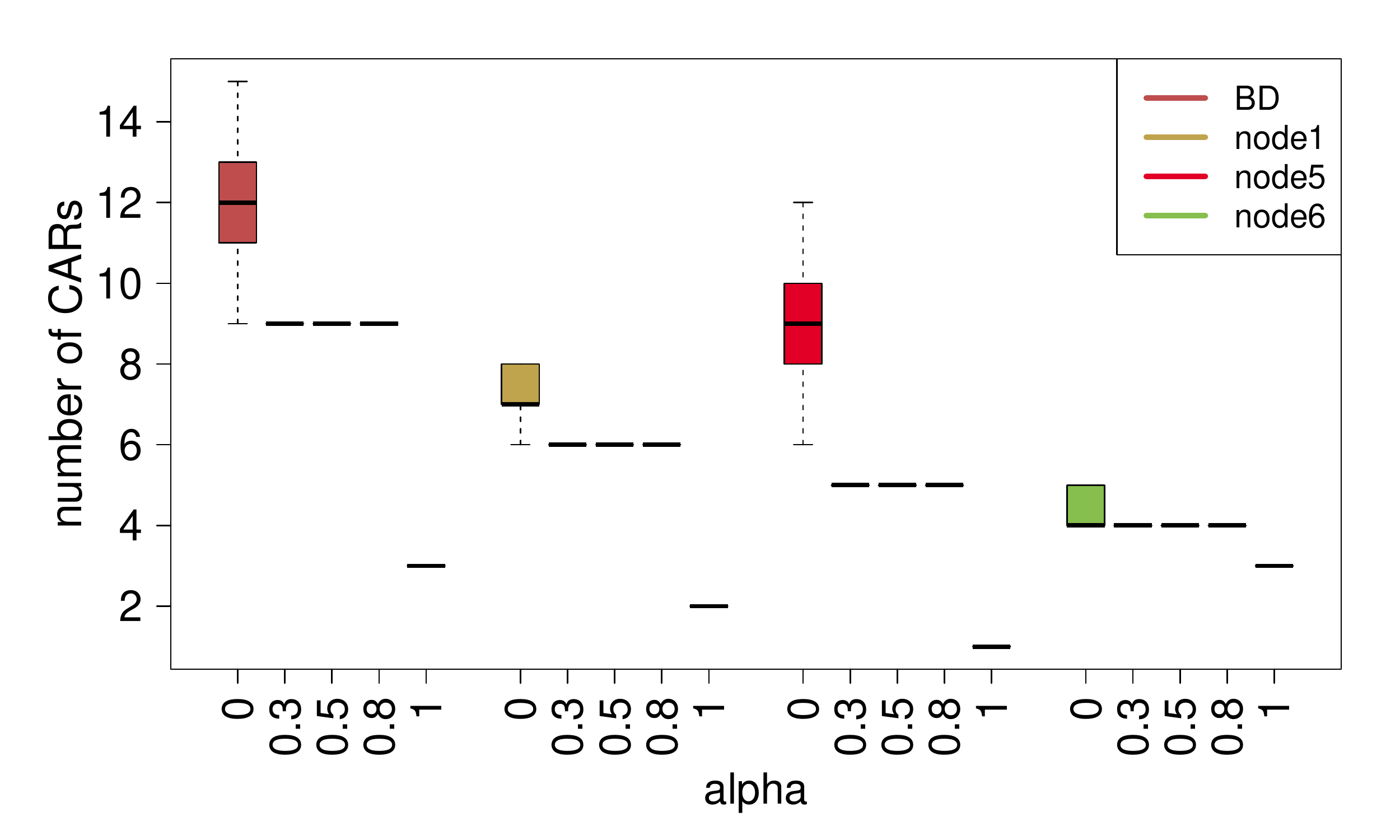}
\caption{\textit{Yersinia} dataset: Reconstructed number of CARs with DeClone weights at $kT=0.1$.}
\label{fig:DeClone_frag01}
\end{figure}

%
%
%
%
\newpage

\pagebreak[4]






\newenvironment{proofsketch}{%
  \renewcommand{\proofname}{Proof sketch}\proof}{\endproof}



\newcommand{\gains}{gains}

\newcommand{\apriori}{\emph{a priori}\xspace}
\newcommand{\prob}[1]{\textsc{#1}\xspace}
\newcommand{\define}[1]{\emph{#1}\xspace}
\newcommand{\taxa}{{\Sigma}} 
\newcommand{\relation}{{\mathcal R}\xspace} 
\newcommand{\genes}{{\Gamma}}

\newcommand{\regions}{{\mathcal C}\xspace} 
\renewcommand{\L}{{\mathcal L}}
\newcommand{\A}{{\mathcal A}}
\newcommand{\I}{{\mathcal I}}
\renewcommand{\S}{{\mathcal S}} 
\renewcommand{\H}{{\mathcal H}} 
\newcommand{\G}{{\mathcal G}}
\newcommand{\Tr}{tr}
\newcommand{\T}{{\mathcal T}}
\newcommand{\C}{{\mathcal C}}
\newcommand{\F}{{\mathcal F}} 
\newcommand{\J}{{\mathcal J}} 
\newcommand{\B}{{\mathcal B}} 
\newcommand{\E}{{\mathcal E}} 
\newcommand{\bH}{\bar{H}\xspace} 
\renewcommand{\O}{{\mathcal O}\xspace}
\renewcommand{\P}{{\mathcal P}}
\newcommand{\R}{{\mathcal R}} 

\renewcommand{\l}{{\ell}}










\subsection*{NP-Hardness for high values of $\alpha$}

In this section we show that {\sf Weighted SCJ labeling} is NP-hard for $33/34 < \alpha < 1$.

\subsubsection*{The Maximum Intersection Matching Problem}

\noindent \textbf{Maximum Intersection Matching Problem (MIMP)} \\
\noindent {\bf Input}: two graphs $G_1$ and $G_2$ with $V(G_1) = V(G_2)$, and an integer $r$.\\
\noindent {\bf Question}: does there exist a perfect matching $M_1$ in $G_1$ 
and a perfect matching $M_2$ in $G_2$ such that $|E(M_1) \cap E(M_2)| \geq r$? \\
 
\begin{figure}[h]
\centering
  \includegraphics[width=1\linewidth]{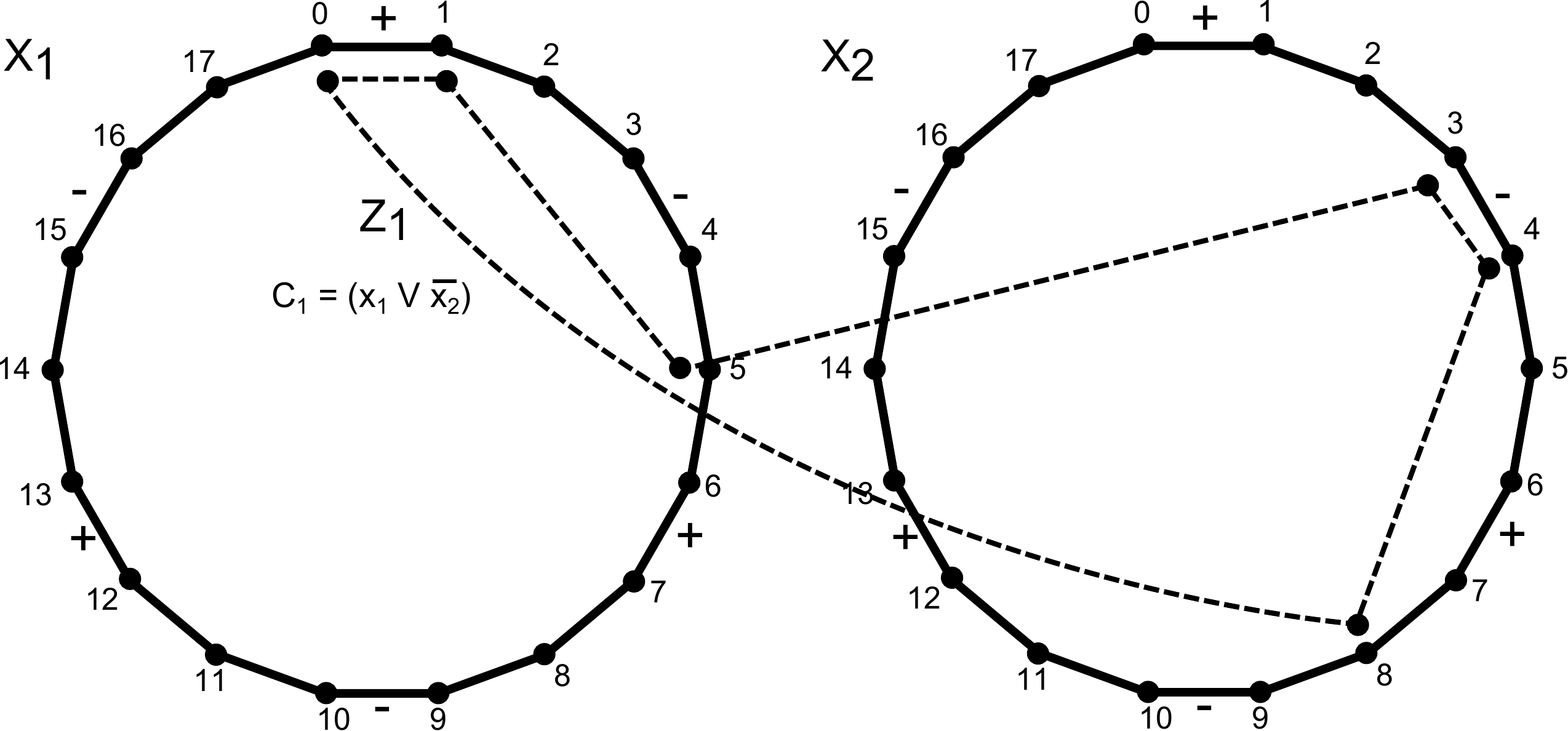}
  \caption{Illustration of a clause gadget used for the \mimp~reduction.  The cycles $X_1$ and $X_2$ are the cycles of $G_1$ corresponding to variables $x_1$ and $x_2$.  The `+' edges are positive, and the `-' edges are negative., The cycle $Z_1$ (dashed edges) corresponds to the clause $C_1 = (x_1 \vee \overline{x_2}$), where here the edges $v^1_0v^1_1$ and $v^2_3v^2_4$ are hypothetically mapped to $C_1$.  Other clause cycles are not shown.}\label{fig:mimp}
\end{figure}

We now introduce the problem used for showing the hardness of \mimp.
The {\sf 3-Balanced-Max-2-SAT} problem is, given $n$ boolean variables $x_1, \ldots, x_n$ 
and a set of $m$ clauses $C = \{C_1, \ldots, C_m\}$, each with  
exactly two literals (e.g. $(x \vee y)$), such that each variable appears in the clauses 
$3$ times positively and $3$ times negatively, to find an assignment 
to the $n$ variables such that a maximum of clauses are satisfied.
Note that $m = 3n$ since each variable appears in $6$ clauses and each clause has $2$ variables.
This problem is NP-hard by a reduction from Max-Cut on $3$-regular graphs~\cite{berman1999some}.
We reduce {\sf 3-Balanced-Max-2-SAT} to \mimp, 
and we begin by describing how we transform a given {\sf 3-Balanced-Max-2-SAT} instance 
into a \mimp~instance.   Figure~\ref{fig:mimp}
illustrates the main ideas behind the reduction.

Let $x_1, \ldots, x_n$ be the set of $n$ variables 
and $C = \{C_1, \ldots, C_m\}$ be the set of clauses
of the given {\sf 3-Balanced-Max-2-SAT} instance.
In our construction the graph $G_1$ corresponds to the variables and $G_2$ to the clauses.

The graph $G_1$ is the disjoint union of $n$ cycles $X_1, \ldots, X_n$
each of length $18$, 
one for each variable $x_i$.
For each $1 \leq i \leq n$, the cycle $X_i$ corresponding to $x_i$ has 
vertices $v^i_0, v^i_1, \ldots, v^i_{17}$ in cyclic order.
Call the edges $v^i_0v^i_1, v^i_6v^i_7$ and $v^i_{12}v^i_{13}$ \emph{positive}, 
and the edges $v^i_3v^i_4, v^i_9v^i_{10}$ and $v^i_{15}v^i_{16}$ \emph{negative}.
Note that $X_i$ has exactly two perfect matchings:
one containing all the positive edges, and one containing all the negative edges.
Since $G_1$ has $n$ such cycles, $G_1$ has $2^n$ perfect matchings, which are in bijection with the possible $x_i$ assignments.
Also note that no two positive or negative edges share a vertex.

Some additional notations are needed before we can construct $G_2$.
Let $C^{+i}_1, C^{+i}_2, C^{+i}_3$ be the $3$ clauses containing variable $x_i$ positively.
Map each of the $3$ positive edges of $X_i$ to a distinct clause arbitrarily 
(e.g. we map $v^i_0v^i_1$ to $C^{+i}_1$, $v^i_6v^i_7$ to $C^{+i}_2$ and $v^i_{12}v^i_{13}$ to $C^{+i}_3$).
Similarly let $C^{-i}_1, C^{-i}_2, C^{-i}_{3}$ be the clauses in which $x_i$ appears negatively, 
and map each negative edge of $X_i$ to one of these clauses in a distinct manner.

Now, $G_2$ has the same vertex set as $G_1$, and is the union of $m$ cycles $Z_1, \ldots, Z_m$ of length $6$, one for each clause in $C$.
Let $x_j$ and $x_k$ be the two variables contained in a clause $C_i$.
Let $v^j_{a}v^j_{a + 1}$ be the edge of the cycle $X_j$ mapped to $C_i$, 
and $v^k_{b}v^k_{b + 1}$ be the edge of the cycle $X_k$ mapped to $C_i$.
Then the cycle $Z_i$ corresponding to $C_i$ is, in cyclic order,  
$v^j_{a}v^j_{a + 1}v^j_{a + 5}v^k_{b}v^k_{b + 1}v^k_{b + 5}v^j_{a}$,
where addition is to be taken modulo $18$.  Thus $Z_i$ contains exactly two edges in common with $G_1$:
$v^j_av^j_{a + 1}$ is from $X_j$ and $v^k_{b}v^k_{b + 1}$ is from $X_k$.
Importantly, $Z_i$ has $2$ possible perfect matchings: one contains
$v^j_{a}v^j_{a + 1}$ and the other has $v^k_{b}v^k_{b + 1}$. 

Intuitively speaking, if an edge $e$ is in both $G_1$ and $G_2$, say on the cycles
$X_i$ of $G_1$ and $Z_j$ of $G_2$,
then $e$ corresponds to the fact that $x_i$ appears in clause $C_j$.
If $e$ is positive, setting $x_i = True$ satisfies $C_j$, and if $e$ is negative, 
setting $x_i = False$ satisfies $C_j$.

We first prove some structural properties of $G_1$ and $G_2$ which will be of use later on.
We denote by $\overline{G}$ the \emph{complement} of a graph $G$.

\begin{lemma}\label{lem:structg1g2}
The graphs $G_1$ and $G_2$ satisfy the following conditions:

\begin{enumerate}
\item 
$|V(G_1)| = |V(G_2)|$ are even;

\item
$G_1$ is the disjoint union of cycles of length at most $18$;

\item
$G_2$ is the disjoint union of cycles of length at most $6$;

\item
$\overline{G_1}$ and $\overline{G_2}$ have a perfect matching in common, i.e. 
there is a set $E \subseteq E(\overline{G_1}) \cap E(\overline{G_2})$
such that $E$ forms a perfect matching in both
$\overline{G_1}$ and $\overline{G_2}$.

\end{enumerate}

\end{lemma}

\begin{IEEEproof}
Conditions 1) and 2) are immediate from our construction.
For 3), we need to show that the $m$ cycles of $G_2$ are disjoint.
We argue that 
each vertex of $G_2$ appears in exactly one cycle. 
First, each vertex $v^j_a$ gets included in at least one cycle $Z_i$: if $v^j_a$ is an endpoint of 
a positive or a negative edge $e$, then $e$ is mapped to some clause $C_i$, and so $v^j_a$ is on the $Z_i$ cycle.
Otherwise, $v^j_a$ is not on a positive or negative edge, 
and thus we have $a \in \{2,5,8,11,14,17\}$.
Therefore $a - 5 \in \{15, 0, 3, 6, 9,12\}$ (modulo $18$).  In this case, $v^j_{a - 5}$ is on a positive or negative edge, and one can check that
$v^j_a$ is in the same cycle $Z_i$ as $v^j_{a - 5}$.
To see that no vertex gets included in two cycles, 
observe that $|V(G_2)| = |V(G_1)| = 18n = 18(m/3) = 6m$.  Thus $G_2$ has $m$ cycles of length $6$ spanning $6m$ vertices, 
which is only possible if no two cycles of $G_2$ intersect. 
For condition 4), there are many choices, and we exhibit one.
Let $1 \leq j \leq n$ and $1 \leq a \leq 18$.  
One can check that in both $G_1$ and $G_2$, $v^j_a$ is not a neighbor of $v^j_{a + 2}$.
Thus the set $E = \{v^j_av^j_{a+2}:1 \leq j \leq n, 0 \leq a \leq 17\}$ (addition modulo $18$)
is a perfect matching of $\overline{G_1}$ and $\overline{G_2}$.
\end{IEEEproof}

Observe that since $G_2$ is a disjoint union of cycles, 
it has $2^m$ perfect matchings since each cycle has $2$ possible perfect matchings.  
We are now ready to prove the hardness of \mimp.


\begin{IEEEproof}[Proof of Theorem~\ref{theorem:MIMP}]
It is easy to see that \mimp~is in NP, since checking that two given matchings 
are perfect and share $r$ edges in the two graphs can be done in polynomial time.
Also, $G_1$ and $G_2$ can obviously be constructed in polynomial time.
To prove NP-hardness, 
we claim that, using the construction described above, at least $r$ clauses of $C$ are satisfiable if and only if 
 $G_1$ and $G_2$ admit a perfect matching with at least 
 $r$ edges in common.
 
($\Rightarrow$) Let $A$ be an assignment of $x_1, \ldots, x_n$ satisfying $r$ clauses of $C$.
In $G_1$, for each $X_i$ cycle, take the positive perfect matching of $X_i$ if 
$x_i$ is positive in $A$, and the negative perfect matching otherwise.
The result is a perfect matching $M_1$ of $G_1$.
To construct a perfect matching $M_2$ of $G_2$, 
let $C_i$ be a clause satisfied by the value of a variable $x_j$ in $A$.
Then, take in $M_2$ the perfect matching of 
$Z_i$ that contains the edge $v^j_av^j_{a+1}$ of $X_j$ mapped to 
$C_i$.  This edge is shared by both our choices in $G_1$ and $G_2$.  We apply this choice of $Z_i$ perfect matching 
for each clause $C_i$ satisfied by $A$ (and pick the matching of $Z_{i'}$ arbitrarily 
for an unsatisfied clause $C_{i'}$).  Since $r$ clauses are satisfied, there must be $r$ common edges 
in $M_1$ and $M_2$.

($\Leftarrow$) Let $M_1$ and $M_2$ be the perfect matchings of $G_1$ and $G_2$
with common edges $F = E(M_1) \cap E(M_2)$, $|F| \geq r$.
Observe that each edge of $F$ must be either positive or negative in $G_1$.
We claim that the following assignment
$A$ satisfies at least $r$ clauses: 
set $x_i = True$ if $F$ has a positive edge in $X_i$, 
set $x_i = False$ if $F$ has a negative edge in $X_i$, 
and set $x_i$ arbitrarily if $F$ has no positive or negative edge in $X_i$.
No $x_i$ can be assigned two values since $F \subseteq M_1$ and $M_1$ is in a perfect matching of $G_1$
(and thus cannot contain both positive and negative edges from the same cycle $X_i$).
Let $Z = \{Z_{i_1}, Z_{i_2}, \ldots, Z_{i_k}\}$ be the clause 
cycles of $G_2$ containing an edge of $F$.
For each $Z_i$ in $G_2$, there can only be one edge
of $Z_i$ in $F$ (recall that each possible perfect matching of $Z_i$ has exactly one edge in common with $G_1$).  
Hence since $|F| \geq r$, 
there must be at least $r$ clause cycles in $Z$, and thus at least $r$ satisfied clauses.
\end{IEEEproof}

\subsubsection*{Hardness of {\sf Weighted-SCJ-labeling}}

Given an \mimp~instance $G_1$ and $G_2$ that satisfies the conditions of Lemma~\ref{lem:structg1g2}, we construct a 
{\sf Weighted-SCJ-labeling} instance consisting of a phylogeny $T$, consistent extant adjacencies
on the leaves of $T$ and weighted ancestral adjacencies on its internal nodes.
Let $2n = |V(G_1)| = |V(G_2)|$.
The {\sf Weighted-SCJ-labeling} instance has $n$ markers, 
and the phylogeny is a tree $T$ with $4$ leaves 
$A, B, C, D$ and 
$3$ internal nodes $X, Y$ and $Z$.  The node $X$ is the parent of $A$ and $B$, the node $Y$ is the parent of $C$ and $D$ and the node $Z$ is the parent of $X$ and $Y$.
Let $E \subseteq E(\overline{G_1}) \cap E(\overline{G_2})$
such that $E$ forms a perfect matching in both
$\overline{G_1}$ and $\overline{G_2}$.
The marker extremities are $V(G_1) = V(G_2)$ and the marker head-tail pairs 
are given by $E$ (i.e. if $xy \in E$, then $x$ and $y$ are the two extremities of a marker).

Since $G_1$ is a union of disjoint cycles of even length, its edges can be partitioned into two disjoint perfect matchings 
$M_{1,1}$ and $M_{1,2}$.  More precisely, let $M_{1,1}$ be a perfect matching of $G_1$.  By removing the set of edges $M_{1,1}$
from $G_1$, we obtain another perfect matching $M_{1,2}$.
The extant adjacencies for the leaf $A$ are given by $M_{1,1}$, and those of $B$ by $M_{1,2}$.
Similarly, $G_2$ can be split into two disjoint perfect matchings $M_{2,1}$ and $M_{2,2}$.
The extant adjacencies of leaf $C$ are given by $M_{2,1}$ and those of leaf $D$ by $M_{2,2}$.


The set of adjacencies under consideration is $\mathcal{A} := E(G_1) \cup E(G_2)$. 
In $X$ (respectively $Y$),
all adjacencies corresponding to an edge of $G_1$ (respectively $G_2$) have weight $1$.  All other adjacencies, 
including all adjacencies in $Z$, have weight $0$.
Note that in this construction, any ancestral adjacency $a$ with non-zero weight 
can be found in some leaf of $T$, which corresponds to the initialization in our method.

Also note that no adjacency connects two extremities
from the same marker.  Indeed, all adjacencies on $T$,
extant or ancestral, are taken from either $G_1$ or $G_2$, whereas marker extremities are 
defined by $E$, which is taken from the complement of $G_1$ and $G_2$.
Thus there is no marker/adjacency inconsistency.

We show that if $\alpha > 33/34$, {\sf Weighted SCJ labeling}
is NP-hard.  This is mainly because we are ``forced'' to assign a perfect matching to 
$X$ and $Y$, as we show in the next Lemma.
We then need to choose these matchings so as to maximize their intersection, 
since this minimizes the SCJ distance on the $XZ$ and $YZ$ branches.

\begin{lemma}\label{lem:perfmatch}
Let $\alpha > 33/34$.
Let $\lambda$ be any labeling for the instance $T$ constructed from $G_1$ and 
$G_2$ as above. 
Then there is a labeling $\lambda'$ 
such that the edges corresponding to $\lambda'(X)$
(respectively $\lambda'(Y)$) 
form a perfect matching of $G_1$ (respectively $G_2$), and 
 $D(\lambda',T) \leq D(\lambda,T)$.
\end{lemma}

\begin{IEEEproof}
Call a cycle $\mathcal{C}$ of $G_1$
(resp. $G_2$) \emph{covered by $\lambda$}
if $\lambda(X)$ (resp. $\lambda(Y)$) has exactly 
$|V(\mathcal{C})|/2$ edges of $\mathcal{C}$, 
i.e. $\lambda(X)$ (resp. $\lambda(Y)$) contains a perfect matching of $\mathcal{C}$.  
If every cycle of $G_1$ and every cycle of $G_2$
is covered by $\lambda$, then the Lemma holds.
Otherwise, we show that if there is an uncovered cycle $\mathcal{C}$, it is advantageous to modify $\lambda$ in order to make $\mathcal{C}$ covered.

Thus let $\mathcal{C}$ be a cycle 
of either $G_1$ or $G_2$ that is not covered by $\lambda$, and denote $c := |V(\mathcal{C})|$.  We have $c \leq 18$.
Let $W = X$ if $\mathcal{C}$ is in $G_1$, and $W = Y$ 
if $\mathcal{C}$ is in $G_2$ instead.
Let $M$ be the edges of $\mathcal{C}$ that correspond to an adjacency labeled by $\lambda(W)$.
Let $d$ be the number of vertices of $\mathcal{C}$ that do not belong to an edge of $M$ 
(i.e. vertices of $\mathcal{C}$ that do not belong to any adjacency of $\lambda(W)$, or that belong to an adjacency that does not correspond to an edge of $\mathcal{C}$).  Since $\mathcal{C}$ is not covered we assume that $d \geq 1$.  
Now, $M$ has $(c - d)/2$ edges.  
As $c$ is even, this implies $d \geq 2$ as it must also be even.
Obtain $\lambda'$ from $\lambda$ 
by first removing all adjacencies of $\lambda(W)$ including a vertex of $\mathcal{C}$,
then adding to $\lambda(W)$ the perfect matching of $\mathcal{C}$ that has the most edges common to $M$, observing that we 
retain at least $|M|/2$ edges of $M$ (and hence remove at most $|M|/2$).
Then the number of adjacencies in $\lambda$ but not in $\lambda'$ 
is at most $d + |M|/2 = d + \frac{(c - d)/2}{2} = d + \frac{c - d}{4} \leq d + \frac{18 - d}{4}$.  
On the other hand at most $c/2 - |M|/2 = c/2 - (c - d)/4 = \frac{c + d}{4} \leq \frac{18 + d}{4}$ adjacencies 
are assigned by $\lambda'$ but not by $\lambda$.
As $X$ and $Y$ have $3$ incident edges, the increase in the SCJ cost incurred 
by changing from $\lambda$ to $\lambda'$ is at most 
$(1 - \alpha) \cdot 3(d + \frac{18 - d}{4} + \frac{18 + d}{4}) = (1 - \alpha)(3d + 27)$.
On the other hand, the cost of adjacency weights is decreased by $\alpha d/2$, since in $\mathcal{C}$, $|M|=(c-d)/2$ edges of weight $1$ are removed, and $c/2$ edges of weight $1$ are added, and so $d/2$ new adjacencies of weight $1$ are gained.
The total difference in cost from $\lambda$ to $\lambda'$ is at most 
$(1 - \alpha)(3d + 27) - \alpha d/2$.
One can check that for $d >= 2$ and $\alpha \geq 33/34$, this total difference is no more than $0$.
Therefore, we can modify $\lambda$ to cover $\mathcal{C}$
at an equal or better cost.  By applying this idea on every cycle of $G_1$ and $G_2$, we get a labeling $\lambda'$ such that 
$\lambda'(X)$ is a perfect matching of $G_1$ and 
$\lambda'(Y)$ is a perfect matching of $G_2$.
\end{IEEEproof}

We can finally show the hardness of {\sf Weighted-SCJ-labeling}.


\begin{IEEEproof}[Proof of Theorem~\ref{theorem:SCJ}]
To prove NP-hardness of {\sf Weighted-SCJ-labeling}, we use the construction described above, which can easily be seen to be feasible in 
polynomial time.
We show that $G_1$ and $G_2$ admit a perfect matching with $k$ common edges 
if and only if $T$ as constructed above admits a labeling
total cost at most $(1 - \alpha)(6n - 2k) + \alpha \cdot 2n$ if $\alpha > 33/34$.

($\Rightarrow$) Let $M_1$ and $M_2$ be perfect matchings of $G_1$ and $G_2$, respectively, 
that share edge set $I = E(M_1) \cap E(M_2)$, with $|I| \geq k$.  We label the internal node $X$ of $T$ with the adjacencies corresponding to $M_1$,
the internal node $Y$ with the adjacencies corresponding to $M_2$,
and internal node $Z$ with the adjacencies corresponding to $I$.
Now, each of the $n$ adjacencies in $\lambda(X)$ is found in either $\lambda(A)$ or $\lambda(B)$ (but not both).
Thus $|\lambda(X) \setminus \lambda(A)| + |\lambda(X) \setminus \lambda(B)| = n$.
Conversely $|\lambda(A) \setminus \lambda(X)| + |\lambda(B) \setminus \lambda(X)| = n$ because half of $\lambda(A) \cup \lambda(B)$ is in $\lambda(X)$.
Thus the $AX$ and $BX$ branches, together, induce an SCJ cost of 
$2n$.  
Similarly, the $CY$ and $DY$ branches induce an SCJ cost of $2n$.
There are $n - k$ adjacencies of $\lambda(X)$ not in $\lambda(Z)$ and all adjacencies of $\lambda(Z)$ are in $\lambda(X)$, 
and so the $XZ$ branch has SCJ cost $n - k$.  Likewise, $YZ$ has SCJ cost $n - k$.
The total SCJ cost is $6n - 2k$.

As for the cost of adjacency weights, $X$ has $2n$ adjacencies of weight $1$,
and $n$ of them are used by $\lambda(X)$.
The same holds for $Y$.  Thus 
$X$ and $Y$, together, induce an ancestral adjacency cost of $4n - 2n = 2n$.
The total weighted cost is $(1 - \alpha)(6n - 2k) + \alpha\cdot 2n$ as desired.

($\Leftarrow$) Suppose that $G_1$ and $G_2$ is a ``no'' instance, \ie no pair of perfect matching 
have $k$ edges in common.  We show that any labeling has cost strictly greater 
than $(1 - \alpha)(6n - 2k) + \alpha\cdot 2n$.
As per Lemma~\ref{lem:perfmatch}, there is an optimal labeling $\lambda$ for $T$ 
such that $\lambda(X)$ is a perfect matching of $G_1$
and $\lambda(Y)$ is a perfect matching of $G_2$.
Choose such an optimal $\lambda$ 
that labels $Z$ with a minimum number of adjacencies.
By the same argument as above, the branches $AX, BX, CY$ and $DY$ incur, together, an SCJ cost of $4n$.
As for $XZ$ and $YZ$, let $a$ be an adjacency in $\lambda(Z)$, and recall that $a$ has weight $0$ in $Z$.  
Then $a$ must be in one of $\lambda(X)$ or $\lambda(Y)$, 
since if $a$ is in neither of $\lambda(X)$ nor $\lambda(Y)$, then it should be removed
as $\lambda$ is suboptimal.  
If $a$ is in $\lambda(X)$ but not in $\lambda(Y)$, presence or absence of $a$ in $\lambda(Z)$
has the same cost, and thus $a\not\in\lambda(Z)$ by our choice of $\lambda$.
Thus $\lambda(Z) \subseteq \lambda(X) \cap \lambda(Y)$.
Since $G_1$ and $G_2$ form a ``no'' instance and $X$ and $Y$ are labeled by a perfect matching, we must have $|\lambda(X) \cap \lambda(Y)| < k$, and therefore $|\lambda(Z)| < k$.  
The cost incurred 
by the $XZ$ and $YZ$ branches is $2n - 2|\lambda(Z)| > 2n - 2k$.
We get that the total SCJ cost incurred by $\lambda$ is strictly greater than $6n - 2k$.
Exactly $2n$ of the $4n$ putative ancestral adjacencies for $X$ and $Y$ are taken by $\lambda$, 
and so the ancestral adjacency cost 
is $2n$.  In total, the cost of the labeling $\lambda$ is strictly higher than $(1 - \alpha)(6n - 2k) + \alpha(2n)$, 
and $T$ is also a ``no'' instance.
\end{IEEEproof}

The above analysis, in particular Lemma~\ref{lem:perfmatch}, can certainly be improved in order to yield better bounds for $\alpha$.
Nevertheless, the question of whether {\sf Weighted SCJ labeling} is NP-hard for every $0 < \alpha < 1$ is wide open.




\end{document}